\documentclass[preprint,notoc]{JHEP3}
\usepackage{epsfig}

\def\to{\rightarrow}

\def\bi{\begin{itemize}}
\def\ei{\end{itemize}}
\def\te{\tilde e}
\def\c1p{C1^\prime}
\def\ta{\tilde a}

\def\tu{\tilde u}

\def\ta{\tilde a}

\def\tb{\tilde b}

\def\tst{\tilde t}

\def\tg{\tilde g}

\def\tw{\widetilde W}
\def\tz{\widetilde Z}

\def\alt{\lesssim}
\def\agt{\gtrsim}
\def\be{\begin{equation}}  
\def\ee{\end{equation}}  
\def\bea{\begin{eqnarray}}  
\def\eea{\end{eqnarray}}  

\newcommand\njp[3]{{\it New\ J.\ Phys.\ }{\bf #1} (#2) #3}

\newcommand\sjp[3]{{\it Sov.\ J.\ Nucl.\ }{\bf #1} (#2) #3}

\def\Isajet{{\sc Isajet}}

\title{Mixed axion/neutralino cold dark matter\\
in supersymmetric models}

\author{Howard Baer$^{a}$, Andre Lessa$^{a}$, Shibi Rajagopalan$^{a,b}$ and 
Warintorn Sreethawong$^a$\\
$^a$Dept.\ of Physics and Astronomy, University of Oklahoma, Norman, OK 73019, USA\\
$^b$Laboratoire de Physique Subatomique et de Cosmologie, UJF Grenoble 1, 
CNRS/IN2P3, INPG, 53 Avenue des Martyrs, F-38026 Grenoble, France\\
E-mail: \email{baer@nhn.ou.edu}, \email{lessa@nhn.ou.edu}, 
\email{shibi@nhn.ou.edu},\email{wstan@nhn.ou.edu}}

\abstract{
We consider supersymmetric (SUSY) models wherein the strong CP problem is solved by 
the Peccei-Quinn (PQ) mechanism with a concommitant axion/axino supermultiplet.
We examine $R$-parity conserving models where the neutralino is the lightest SUSY particle, 
so that a mixture of neutralinos $and$ axions serve as cold dark matter ($a\tz_1$ CDM).
The mixed $a\tz_1$ CDM scenario can match the measured dark matter abundance for
SUSY models which typically give too low a value of the usual thermal neutralino abundance, 
such as models with wino-like or higgsino-like dark matter.
The usual thermal neutralino abundance can be greatly enhanced by the decay of 
thermally-produced axinos ($\ta$) to neutralinos, followed by neutralino re-annihilation at 
temperatures much lower than freeze-out. In this case, the relic density is usually
neutralino dominated, and goes as $\sim (f_a/N)/m_{\ta}^{3/2}$.
If axino decay occurs before neutralino freeze-out, then instead the
neutralino abundance can be augmented by relic axions to match the measured
abundance.
Entropy production from late-time axino decays can diminish the axion abundance, but
ultimately not the neutralino abundance.
In $a\tz_1$ CDM models, it may be possible to detect both a WIMP and an axion as 
dark matter relics.
We also discuss possible modifications of our results due to production and
decay of saxions.
In the appendices, we present expressions for the Hubble expansion rate and the 
axion and neutralino relic densities in
radiation, matter and decaying-particle dominated universes.
}

\begin{document}

%========================================================================================
\section{Introduction}
\label{sec:intro}
%========================================================================================

The Standard Model of particle physics is beset by two major fine-tuning problems.
The first occurs in the Higgs sector of the electroweak model, 
where quantum corrections to the Higgs mass push $m_h$ up to the energy scale associated 
with the cut-off of the theory, where new physics is expected to enter. 
The electroweak fine-tuning problem is elegantly solved by
the introduction of weak scale supersymmetry (SUSY)\cite{wss}. A consequence of 
weak scale SUSY is that
supersymmetric matter states should exist at or around the electroweak scale, and ought to be
detectable at the CERN LHC\cite{bblt}.

The second fine-tuning problem arises in the QCD sector. Here, t'Hooft's solution to the
$U(1)_A$ problem\cite{thooft} via instantons and the $\theta$ vacuum requires the existence of a 
$CP$ violating term 
\be 
{\cal L}_\theta = \frac{\theta}{32\pi^2}F_{A\mu\nu}{\tilde F}_A^{\mu\nu}
\ee 
in the QCD Lagrangian\cite{axreviews}.
A second contribution to ${\cal L}_\theta$ arises from the electroweak sector, and is proportional
to $arg(det{\cal M})$, where $\cal M$ is the quark mass matrix\cite{arg}. 
Measurements of the neutron EDM tell us that the combination $\theta +arg(det{\cal M})\equiv \bar{\theta}$ 
must be $\alt 10^{-11}$\cite{nedm}. 
Explaining why $\bar{\theta}$ is so small is known as the strong $CP$ problem.
The strong $CP$ problem is elegantly solved by the Peccei-Quinn mechanism\cite{pq} 
and its concommitant {\it axion} $a$\cite{ww}.
Models of an ``invisible axion'' with $PQ$ breaking scale $f_a/N \agt 10^9$ GeV 
(with $N$ being the color anomaly factor, which is 1 for KSZV\cite{ksvz} models 
and 6 for DFSZ\cite{dfsz} models) 
allow for a solution to the strong $CP$ problem while eluding astrophysical constraints 
arising due to energy loss from stars in the form of axion radiation\cite{astro}. 

Of course, the SUSY solution to the electroweak fine-tuning, and the PQ solution to the strong
fine-tuning are not mutually exclusive. In fact, each complements the other\cite{pqsusy}, 
and both are expected to
arise rather naturally from superstring models\cite{witten}. In models which invoke both
$R$-parity conserving SUSY and the PQ solution to the strong $CP$ problem, the dark matter of the
universe is expected to consist of a {\it mixture} of both the axion and the lightest-SUSY-particle (LSP).
Many previous studies have focused on the possibility of an axino $\ta$ as the LSP\cite{rtw,ckkr,axinorev}, 
giving rise to mixed axion/axino cold dark matter (CDM)\cite{axinoDM}. 
In this paper, we explore instead the possibility that the 
lightest neutralino $\tz_1$ is the LSP, thus giving rise to mixed axion/neutralino ($a\tz_1$) CDM.
In the case of mixed $a\tz_1$ CDM, it may be possible to detect relic axions as well as
relic neutralinos (as WIMPs).

The case of neutralino CDM in the PQ-augmented MSSM has been considered previously by Choi {\it et al.}\cite{ckls}.
In Ref. \cite{ckls}, the authors considered the case of neutralino dark matter where 
$m_{\ta}>m_{\tz_1}$. They presented approximate expressions to estimate the relic density of
neutralinos $\Omega_{\tz_1}h^2$. They found that neutralinos can be produced thermally as usual, 
but also that their abundance can be {\it augmented} by thermal production of axinos in the early universe, 
followed by axino cascade decays into the stable $\tz_1$ state. 
However, the neutralino abundance could also be
{\it diminished} by two effects. 
The first is that even after neutralino freeze-out, the additional late-time injection of
neutralinos into the cosmic soup from axino decay can cause a re-annihilation effect. The second diminution effect 
occurs when late decaying axinos inject entropy into the early universe after neutralino freeze-out, 
thus possibly diluting the frozen-out neutralino abundance.

In this paper, our goal is to present explicit numerical calculations of the relic abundance 
of mixed $a\tz_1$ CDM in SUSY models. 
To this end, we include several new effects. 
\bi
\item First, we note that in the PQMSSM with a $\tz_1$ as LSP, 
the dark matter will consist of an axion/neutralino
admixture, so we always account for the axion contribution to the total DM abundance. 
\item Second, we account for the measured abundance of CDM as is given by the recent 
WMAP7 analysis\cite{wmap7}:
\be
\Omega_{\rm DM}h^2=0.1123\pm 0.0035\ \ \ {\rm at\ 68\%\ CL} .
\ee
We seek to establish under what conditions of model parameters the theoretical prediction for
the relic abundance of mixed $a\tz_1$ CDM can be in accord with the measured value.
\item Third, we seek to establish whether, when fulfilling the WMAP measured abundance, the mixed
$a\tz_1$ DM is dominantly axion or dominantly neutralino, or a comparable mixture.
Such an evaluation is important for determining the relative prospects of axion 
and WIMP direct detection experiments.
\ei

The remainder of this paper is organized as follows. In Sec. \ref{sec:bm}, we introduce two SUSY benchmark models-- 
the first from the minimal supergravity (mSUGRA) model\cite{msugra} in the hyperbolic branch/focus point region\cite{hb_fp}
(where neutralinos are expected to be of the mixed higgsino variety) 
and the second arising from the gaugino AMSB\cite{amsb} model\cite{shanta,inoamsb} (where the
neutralino is of the nearly pure wino variety). 
These benchmarks will be used for illustrative calculations of mixed $a\tz_1$ CDM. 
In Sec. \ref{sec:decays}, we calculate the axino decay rate into all possible modes, while in Sec. \ref{sec:prod}
we discuss the thermal production of axinos in the early universe. 
In Sec. \ref{sec:axdom}, we compute the temperature $T_D$ at which axinos decay, 
and the temperature $T_e$ at which they might dominate the energy density of the universe.
In Sec. \ref{sec:vacmis} and in Sec. \ref{sec:z1}, we discuss the computation of the axion and the neutralino relic density.
 In the case of neutralino production, we pay special attention to the neutralino re-annihilation which takes place due to injection
of neutralinos into the early universe from axino decay. 
In Sec. \ref{sec:algor} we present our algorithm for estimating the mixed $a\tz_1$ CDM abundance in SUSY models. 
In Sec. \ref{sec:results}, we present our numerical results, and in
Sec. \ref{sec:saxion} we discuss the effect of including saxion production and decay in the calculation.
In Sec. \ref{sec:conclude}, we present our conclusions. In the appendices, we present explicit formulae for
the Hubble expansion rate, the axion abundance, and the thermal neutralino abundance for a universe that is either 
radiation-dominated (RD), matter-dominated (MD), or decaying-particle-dominated (DD).

\section{Two benchmark scenarios}
\label{sec:bm}

Supersymmetric models with mixed $a\tz_1$ CDM will always have a neutralino
dark matter component, which is mainly enhanced by axino production and decay, 
but might also be diminished by further annihilations and by entropy injection. 
In addition, there will always be an axion
component to the CDM, which arises as usual via vacuum misalignment. In SUGRA-based models
with gaugino mass unification, the lightest neutralino $\tz_1$ is usually bino-like, and
annihilation reactions are suppressed by their $p$-wave annihilation cross sections, leading to
an overabundance of neutralinos\cite{axdm}. Since SUSY models with mixed $a\tz_1$ CDM will have to crowd both
$\tz_1$s and axions into the overall relic density, we find that the most promising models
to realize mixed $a\tz_1$ CDM are those leading to an {\it under}abundance of the usual 
calculation of thermally-produced 
neutralinos, such as models with either a higgsino-like or wino-like $\tz_1$ state\cite{shibi}: 
such neutralinos annihilate via $s$-wave rather than $p$-wave reactions,
and naively suffer a {\it dearth} of relic abundance.

We will thus choose two benchmark cases to examine mixed $a\tz_1$ CDM. The first occurs in the
hyperbolic branch/focus point region\cite{hb_fp} of mSUGRA\cite{msugra} and has a mixed 
higgsino-wino-bino $\tz_1$ state, while the second is taken from the 
gaugino-anomaly-mediated-SUSY-breaking model (inoAMSB)\cite{inoamsb}, 
which leads to a nearly pure wino-like $\tz_1$ state. 

%%%%%%%%%%%%%%%%%%%%%%%%%%%%%%%%%%%%%%%%%%%%%%%%%%%%%%%%%%%%%%%%%%%%%%%%%%%%%%%%%%%%%%%
%\TABLE{
\begin{table}\centering
\begin{tabular}{lcc}
\hline
 & BM1  & BM2  \\
 & (HB/FP) & (inoAMSB) \\
\hline
$m_0$ & 4525 & 0 \\
$m_{1/2}$  & 275 & -- \\
$A_0$  & 0  & 0 \\
$\tan\beta$  & 10 & 10 \\
$m_{3/2}$ [TeV] & -- & 50 \\
\hline
$\mu$      & 137.2 & 599.4 \\
$m_{\tg}$ & 810.4 & 1129.7 \\
$m_{\tu_L}$ & 4517.0 & 993.9 \\
$m_{\tst_1}$& 2608.1 & 861.4 \\
$m_{\tb_1}$ & 3687.6 & 926.1 \\
$m_{\te_L}$ & 4520.1 & 229.4 \\
$m_{\tw_1}$ & 121.1 & 142.4 \\
$m_{\tz_4}$ & 273.4 & 616.3 \\ 
$m_{\tz_3}$ & 149.8 & 606.0 \\ 
$m_{\tz_2}$ & 143.1 & 443.6 \\ 
$m_{\tz_1}$ & 87.9 &  142.1 \\ 
$m_A$       & 4458.1 &  633.6 \\
$m_h$       & 119.6 &  112.1 \\ 
\hline
%$\Delta a_\mu$ & $12.5\times 10^{-8}$ & $1.6\times 10^{-9}$ \\
%$BF(b\to s\gamma )$ & $3.1\times 10^{-4}$ & $3.8\times 10^{-4}$ \\
%$BF(B_s\to\mu\mu)$ & $3.8\times 10^{-9}$ & $3.8\times 10^{-9}$  \\
$v_4^{(1)}$ & 0.65 & 0.01  \\
$\Omega^{std}_{\tz_1} h^2$ & 0.05 & 0.0016  \\
$T_{fr}$ & $m_{\tz_1}/23.2$ & $m_{\tz_1}/31.2$ \\
$\langle\sigma v\rangle\ [{\rm GeV}^{-2}]$ &  $4.3\times 10^{-9}$ & 
$1.8\times 10^{-7}$ \\
$\sigma (\tz_1 p)$ [pb] & $3.2\times 10^{-8}$  & $4.3\times 10^{-9}$ \\
\hline
\end{tabular}
\caption{Masses and parameters in~GeV units for the HB/FP and inoAMSB benchmark points. 
computed with \Isajet\,7.81 using $m_t=173.3$ GeV.
}
\label{tab:bm}
\end{table}
%%%%%%%%%%%%%%%%%%%%%%%%%%%%%%%%%%%%%%%%%%%%%%%%%%%%%%%%%%%%%%%%%%%%%%%%%%%%%%%%%%%%%%%

\subsection{Mixed higgsino DM in HB/FP region of mSUGRA model}

The spectra for our first benchmark model, BM1, is generated within the minimal
supergravity, or mSUGRA model, using the Isasugra\cite{isasug}
spectrum generator from Isajet v7.81\cite{isajet}. The input parameters are taken as
\bi
\item $(m_0,\ m_{1/2},\ A_0,\ \tan\beta ,\ sign(\mu )=\ 
(4525\ {\rm GeV},\ 275\ {\rm GeV},\ 0,\ 10,\ +)$
\ei
with $m_t=173.3$ GeV.
The $\tz_1$ has mass 87.9 GeV, while the calculated thermal abundance of neutralinos
from IsaReD\cite{isared} is $\Omega_{\tz_1}^{std}h^2=0.05$. Many sparticle masses
and low energy observables are listed in Table \ref{tab:bm}. Since the weak scale value 
of the superpotential $\mu$ term is only 137.2 GeV, the $\tz_1$ is of mixed
higgsino-bino-wino type.

\subsection{Wino-like DM from gaugino AMSB}

The spectra for benchmark point BM2 of Table \ref{tab:bm} is generated within
the gaugino AMSB model which is expected to arise in string theories with
moduli-stabilization via fluxes and a large volume compactification\cite{shanta}. 
In the inoAMSB model, the parameters are taken as
\bi
\item $m_{3/2},\ \tan\beta ,\ sign( \mu )=\ (50\ {\rm TeV},\ 10,\ +)$
\ei
while $m_0=A_0=0$ at the GUT scale. The GUT scale gaugino masses take the 
AMSB form, and are given at $m_{GUT}$ as $M_1=1065.3$ GeV, 
$M_2=161.4$ GeV and $M_3=-484$ GeV.
This choice leads to a neutralino state $\tz_1$ which is nearly pure wino, with 
a thermal relic abundance of $\Omega_{\tz_1}^{std}h^2=0.0016$ as indicated in Table \ref{tab:bm}.

%========================================================================================
\section{Axino decays in the mixed axion-neutralino CDM scenario}
\label{sec:decays}
%========================================================================================

Since we assume $m_{\ta}>m_{\tz_1}$, an important element in the thermal history of the 
universe with mixed $a\tz_1$ CDM will be knowledge of the axino decay width. 

The axino-gluon-gluino coupling is given by
\be
\mathcal{L}_{\ta\tg g}= i\frac{\alpha_s}{16\pi(f_a/N)}\bar{\ta}\gamma_5[\gamma^\mu,\gamma^\nu]
          \tg_A F_{A\mu\nu} .
\ee
Evaluation of the $\ta\to g\tg$ decay width gives a result in accord with Ref. \cite{ckls}, and is given by
\be
\Gamma(\ta\to\tg  g)=	\frac{8\alpha^2_s}{128\pi^3(f_a/N)^2}m_{\ta}^3\left(1-\frac{m_{\tg}^2}{m_{\ta}^2}\right)^3. \label{eq:Gammag}
\ee

Also, the axino-bino-$B_\mu$ coupling is given by
\be
\mathcal{L}_{\ta{\tilde B} B}= i\frac{\alpha_Y C_{aYY}}{16\pi(f_a/N)}\bar{\ta}\gamma_5
[\gamma^\mu,\gamma^\nu]\tilde B B_{\mu\nu} ,
\ee
where $\alpha_Y=g^{\prime 2}/4\pi$, $B_{\mu\nu}=\partial_\mu B_\nu -\partial_\nu B_\mu$ and
$C_{aYY}$ is a model-dependent factor equal to $8/3$ in the DFSZ model, and $0,\ 2/3,\ 8/3$ for
heavy quark charges $e_Q=0,\ -1/3,\ +2/3$ in the KSVZ model. For numerical purposes, we will
take $C_{aYY}=8/3$ since that case occurs in both models.
We then find: 
\be
\Gamma(\ta\to\tz_i +\gamma)=	\frac{\alpha^2_Y C^2_{aYY}\cos^2\theta_w v_4^{(i)2}}{128\pi^3(f_a/N)^2}
m_{\ta}^3\left(1-\frac{m_{\tz_i}^2}{m_{\ta}^2}\right)^3 
\ee
and
\bea
\Gamma(\ta\to\tz_i +Z)&=& \frac{\alpha^2_Y C^2_{aYY}\sin^2\theta_w v_4^{(i)2}}{128\pi^3(f_a/N)^2}
m_{\ta}^3\lambda^{1/2}\left(1,\frac{m_{\tz_i}^2}{m_{\ta}^2},\frac{m^2_Z}{m_{\ta}^2}\right) \\
& &\cdot\left\{\left(1-\frac{m_{\tz_i}^2}{m_{\ta}^2}\right)^2+3\frac{m_{\tz_i} m^2_Z}{m_{\ta}^3}-
\frac{m^2_Z}{2m_{\ta}^2}\left(1+\frac{m_{\tz_i}^2}{m_{\ta}^2}+\frac{m^2_Z}{m_{\ta}^2}\right)\right\} ,
\eea
where $v_4^{(i)}$ is the bino fraction of $\tz_i$ in the notation of Ref. \cite{wss}.

\begin{figure}[t]
\begin{center}
\includegraphics[angle=-90,width=10cm]{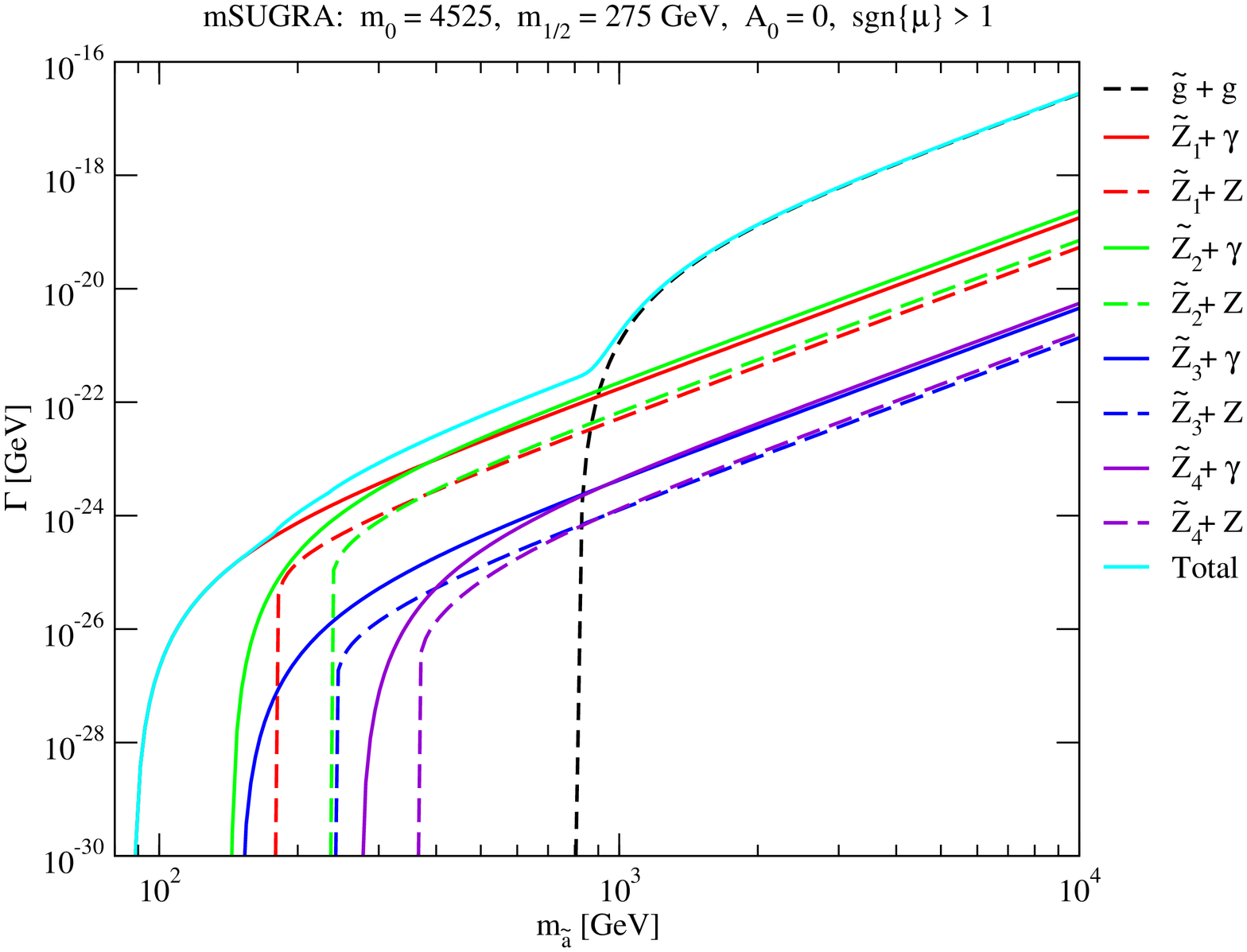} \\[24pt]
\includegraphics[angle=-90,width=10cm]{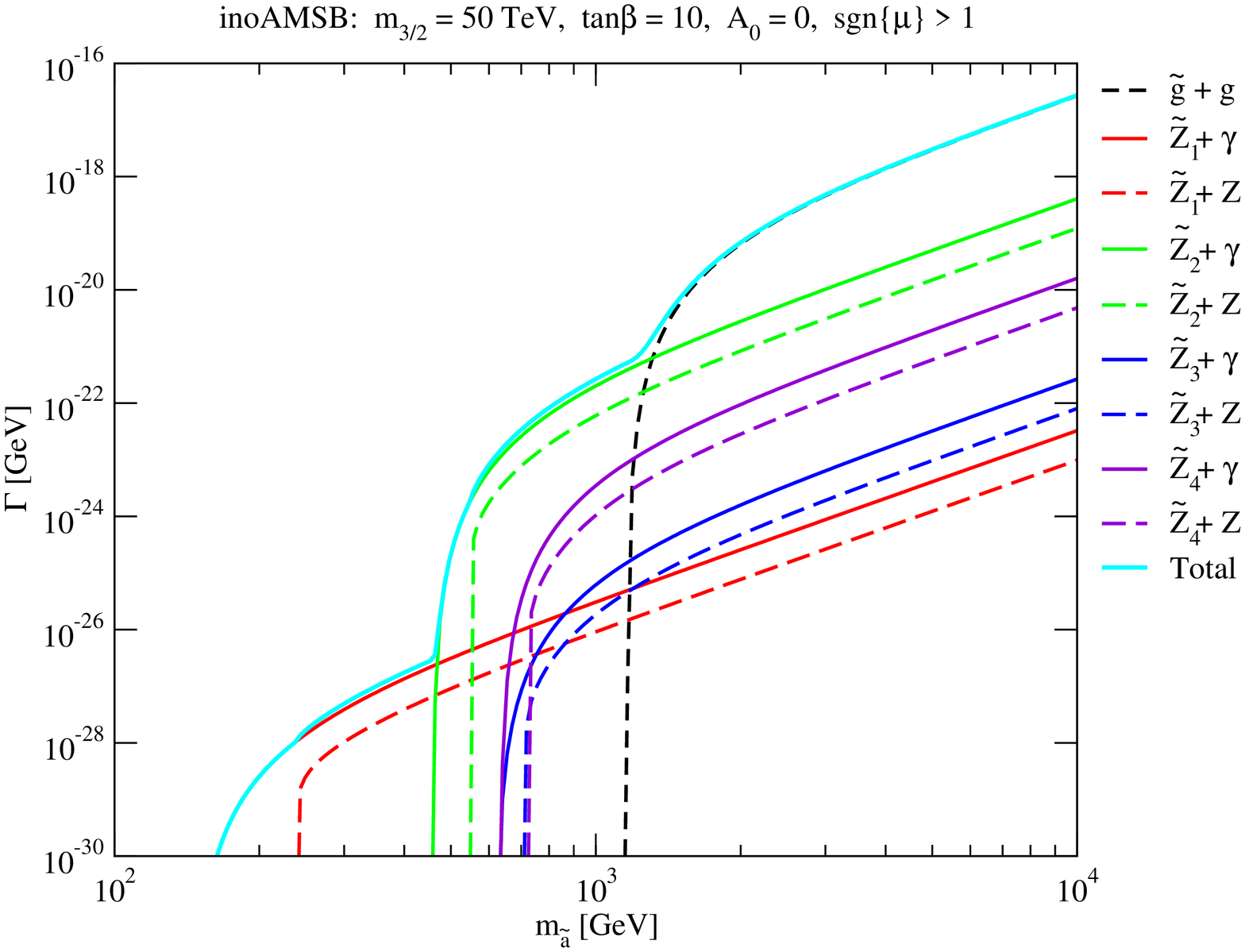}
\caption{Partial and total decay width of axinos versus $m_{\ta}$ for BM1 in the
mSUGRA model with $(m_0,m_{1/2},A_0,\tan\beta,sign(mu)) = (4525,275,0,10,+)$
and for BM2 in the inoAMSB model with $m_{3/2}=50$ TeV, $\tan\beta =10$ and $\mu >0$.
We take $f_a/N=10^{12}$ GeV.
}
\label{fig:axwidth}
\end{center}
\end{figure}

The axino partial and total widths are exhibited in Fig. \ref{fig:axwidth} for
{\it a}). the HB/FP model and {\it b}). the inoAMSB model.
We see that at very low values of $m_{\ta}\sim m_{\tz_1}$, only the decay
$\ta\to \tz_{1} \gamma$ is open, and $\Gamma_{\ta}$ is very small.
As $m_{\ta}$ increases, additional decay modes become allowed, and 
contribute to $\Gamma_{\ta}$. In frame {\it b})., $\Gamma_{\ta}$ enjoys
a large increase when the decay to $\tz_2\gamma$ opens up, since
in the inoAMSB case, the $\tz_2$ is mainly bino-like. Once the
decay to $\tg g$ opens up, this mode is dominant. Decays to $\tz_i Z$ 
are always subdominant to $\tz_i\gamma$, owing to the fact that the photon
has a larger $B_\mu$ component than does the $Z$.
%
%%%%%%%%%%%%%%%%%%%%%%%%%%%%%%%%%%%%%%%%%%%%%%%%%%%%%%%%%%%%%%%%%%%
%%%%%%%%%%%%%%%%%%%%%%%%%%%%%%%%%%%%%%%%%%%%%%%%%%%%%%%%%%%%%%%%%%%

\section{Thermal production of axinos in the early universe}
\label{sec:prod}

If the reheat temperature after inflation, $T_R$, exceeds the axino decoupling temperature\cite{rtw},
\be
T_{dcp}=10^{11}\ {\rm GeV}\left(\frac{f_a/N}{10^{12}\ {\rm GeV}}\right)^2
\left(\frac{0.1}{\alpha_s}\right)^3 ,
\ee
then reheat occurred before decoupling which allowed the axinos to reach 
thermal equilibrium. Their number density at the time of decoupling is given in terms of the 
yield variable, $Y\equiv n/s$, as
\be
Y_{\ta}=\frac{n_{\ta}}{s}|_{T_{dcp}} =\frac{135\zeta (3)}{4\pi^4}\frac{1}{g_* (T_{dcp})} ,
\label{eq:axinoTE}
\ee
with $\zeta (3)\simeq 1.202$, 
$g_*(T_{dcp})$ is the effective number of relativistic degrees of freedom at temperature $T_{dcp}$, and
$s=\frac{2\pi^2}{45}g_{*}T^3$ is the entropy density of radiation\footnote{For simplicity we assume
$g_{*S}(T) = g_*(T)$, where $g_{*S}$ is the total number of relativistic degrees of freedom at $T$ used to compute $s$.}.

In the other case, where $T_R<T_{dcp}$, the axinos were never in thermal equilibrium in the early universe.
However, they could still be produced thermally via radiation off of other particles 
in thermal equilibrium~\cite{ckkr,steffen}. Here, we adopt a recent calculation of the thermally produced
axino yield from Strumia~\cite{strumia}:
\be
Y_{\ta}^{\rm TP}=4.5\times 10^{-9} g_s^4 F(g_s)
\frac{T_R}{10^4\ {\rm GeV}}\left(\frac{10^{11}}{f_a/N}\right)^2 ,
\label{eq:axinoprod}
\ee
with $F(g_s)\sim 20 g_s^2\ln\frac{3}{g_s}$, and $g_s$ is the strong coupling constant
evaluated at $Q=T_R$.

If we suppose that each produced axino will cascade decay into the 
$\tz_1$ state, we can naively estimate the decay-induced neutralino abundance by

%Naively, each produced axino will cascade decay into the $\tz_1$ state, 
%potentially giving rise to a decay induced  neutralino abundance of

%
\be
\Omega_{\tz_1}^{\ta}h^2=\frac{m_{\tz_1}}{m_{\ta}}\Omega_{\ta}^{TP}h^2  .
\label{eq:z1fromaxino}
\ee
This estimate of the neutralino abundance from the decay of thermally-produced 
axinos is shown in Fig. \ref{fig:Oh2_fa} as a function of $f_a/N$ for 
$T_R=10^4$, $10^7$ and $10^{10}$ GeV (green, blue and red curves, respectively),
and for $m_{\tz_1}=50$, 200 and 400 GeV (lower to upper curves). We see that the
neutralino relic density from axino decay can be enormous, and it typically 
dominates over the thermal neutralino abundance as calculated, for instance, from 
the mSUGRA model. From our naive estimate, we see that a tremendous 
overproduction of neutralinos is obtained from thermal axino production and 
decay except when considering parameter regions of high $f_a/N$ or low $T_R$ where 
the axino production rate is suppressed.
%Pushing $f_a/N$ to high values suppresses the effective axino coupling, 
%and  thus suppresses the possible overproduction of neutralinos from axino decay.
%
%%%%%%%%%%%%%%%%%%%%%%%%%%%%%%%%%%%%%%%%%%%%%%%%%%%%%%%%%%%%%%%%%%%
\FIGURE[t]{
\includegraphics[width=10cm]{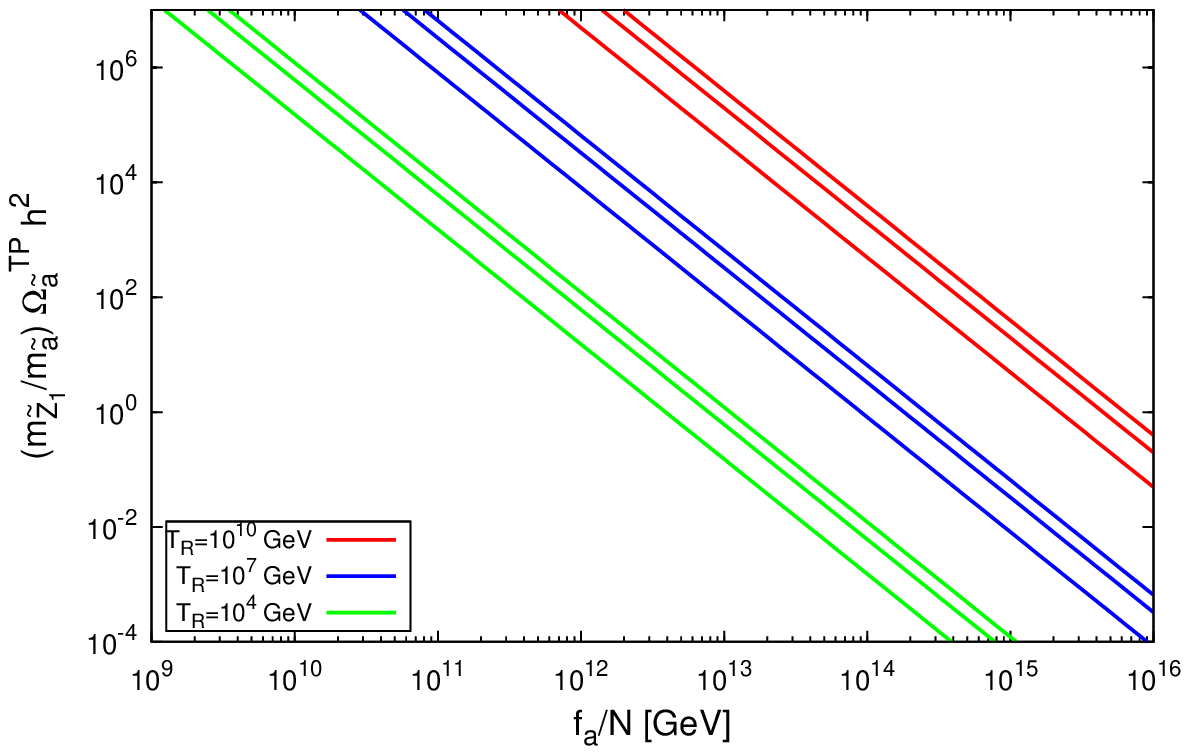}
\caption{Plot of $\Omega_{\tz_1}^{\ta}h^2$ versus $f_a/N$  for various
values of $T_R$ and $m_{\tz_1}=50$, 200 and 400 GeV (lower-to-upper).
}\label{fig:Oh2_fa}}
%%%%%%%%%%%%%%%%%%%%%%%%%%%%%%%%%%%%%%%%%%%%%%%%%%%%%%%%%%%%%%%%%%%

%In Fig. \ref{fig:Oh2_tr}, we show the same plot for neutralino production from axino decay, 
%but this time versus $T_R$, for fixed values of $f_a/N=10^{10}$, $10^{12}$ and $10^{14}$
%GeV (magenta, blue, red, respectively), with $m_{\tz_1}=50$, 200 and 400 GeV in
%ascending order.We see here that, naively, we will obtain a tremendous overproduction of
%neutralinos from thermal axino production and decay unless the axino production rate is
%suppressed by either high $f_a/N$ or low $T_R$.
%
%%%%%%%%%%%%%%%%%%%%%%%%%%%%%%%%%%%%%%%%%%%%%%%%%%%%%%%%%%%%%%%%%%%
%\FIGURE[t]{
%\includegraphics[width=10cm]{omegalsp_adecay_tr.eps}
%\caption{Plot of $\Omega_{\tz_1}^{\ta}h^2$ versus $T_R$  for various
%values of $f_a/N$ and $m_{\tz_1}=50$, 200 and 400 GeV (lower-to-upper).
%}\label{fig:Oh2_tr}}
%%%%%%%%%%%%%%%%%%%%%%%%%%%%%%%%%%%%%%%%%%%%%%%%%%%%%%%%%%%%%%%%%%%

\section{Axino domination of the Universe}
\label{sec:axdom}

Once the decay width of the axino has been calculated as in Sec.
\ref{sec:decays}, we may then calculate the axino lifetime or, alternatively, 
the temperature of radiation at the time scale when nearly all axinos have 
decayed, $T_D$.  This is achieved by equating the Hubble and the decay rates
\be
H(T_D) = \Gamma_{\ta},
\ee
which implies
\be
T_D=\sqrt{\Gamma_{\ta}M_P}/(\pi^2g_*(T_D)/90)^{1/4}, \label{eq:TD}
\ee
where $M_P$ is the reduced Planck mass $M_P\simeq 2.4\times 10^{18}$ GeV.
The temperature $T_D$ also corresponds to the temperature of radiation
when entropy injection from axino decays is nearly finished.  We assume an 
exponential decay law for $\ta$ which implies that the axino decays only cause a 
slow-down in the rate of cooling of the expanding universe\cite{scherrer}.  In 
this case, $T_D$ should not be interpreted as a ``second reheat" temperature.

%While $T_D$ can be considered as
%a second re-heat temperature, it is shown in Ref. \cite{scherrer} that the axino decays
%instead only cause a slow-down in rate of cooling of the expanding universe.

As the universe-- filled with axinos and radiation-- expands and cools, $T$ drops below $m_{\ta}$ 
and the axinos become non-relativistic. At that point, the axino energy density decreases as $T^3$, 
while the radiation energy density continues to decrease faster as $T^4$. At some temperature
$T_e$, the energy density of axinos will dominate the universe
if the axinos have not yet decayed, {\it i.e.} provided $T_e > T_D$. By equating the energy density of radiation
$\rho_R=\pi^2 g_*T^4/30$ with the energy density of axinos $\rho_{\ta}|_{T<m_{\ta}}=m_{\ta}Y_{\ta}s$,
we can determine $T_e$ to be\cite{ckls}
\be
T_e={4\over 3}m_{\ta}Y_{\ta} .
\ee
Furthermore, we can estimate the ratio of entropy after to the entropy before axino decay,
r, again where $T_e>T_D$, by\cite{scherrer}
\be
r=\frac{S_f}{S_0}\simeq \frac{4m_{\ta}Y_{\ta}}{3T_D}=T_e/T_D .
\label{eq:r}
\ee

In Fig. \ref{fig:TD11}{\it a})., we plot the values of $T_D$, $T_e$ and 
the temperature of neutralino freeze-out $T_{fr}$ 
versus $m_{\ta}$ for the benchmark BM1: the HB/FP region of the mSUGRA model.
We take $f_a/N=10^{12}$ GeV, and show $T_e$ for $T_R=10^6$ and $10^{10}$ GeV.
For values of $T_D\alt 2$ MeV (gray shaded region), the parameters would be likely excluded 
because axinos would dump entropy after BBN has started.
For $T_D>T_{fr}$-- the high $m_{\ta}$ region-- 
axinos decay to neutralinos before freeze-out. In this case, the
neutralinos from axino decays thermalize and
the neutralino relic abundance $\Omega_{\tz_1}h^2$
is given as usual by the standard calculation of WIMP thermal abundance ($\Omega_{\tz_1}^{0} h^2$).
The region where $T_e>T_D$ is where axinos can dominate the universe.
This occurs for $m_{\ta}\alt 8$ TeV in the $T_R=10^{10}$ GeV case. Furthermore, 
the ratio between the $T_e$ and $T_D$ curves gives an approximate
measure of the entropy injection from axino decay.
In frame {\it a}). for $T_R=10^6$ GeV, axinos never dominate
the universe, since they decay prior to the point where the axino
energy density exceeds that of radiation.
%%%%%%%%%%%%%%%%%%%%%%%%%%%%%%%%%%%%%%%%%%%%%%%%%%%%%%%%%%%%%%%%%%%
\FIGURE[t]{
\includegraphics[width=10cm]{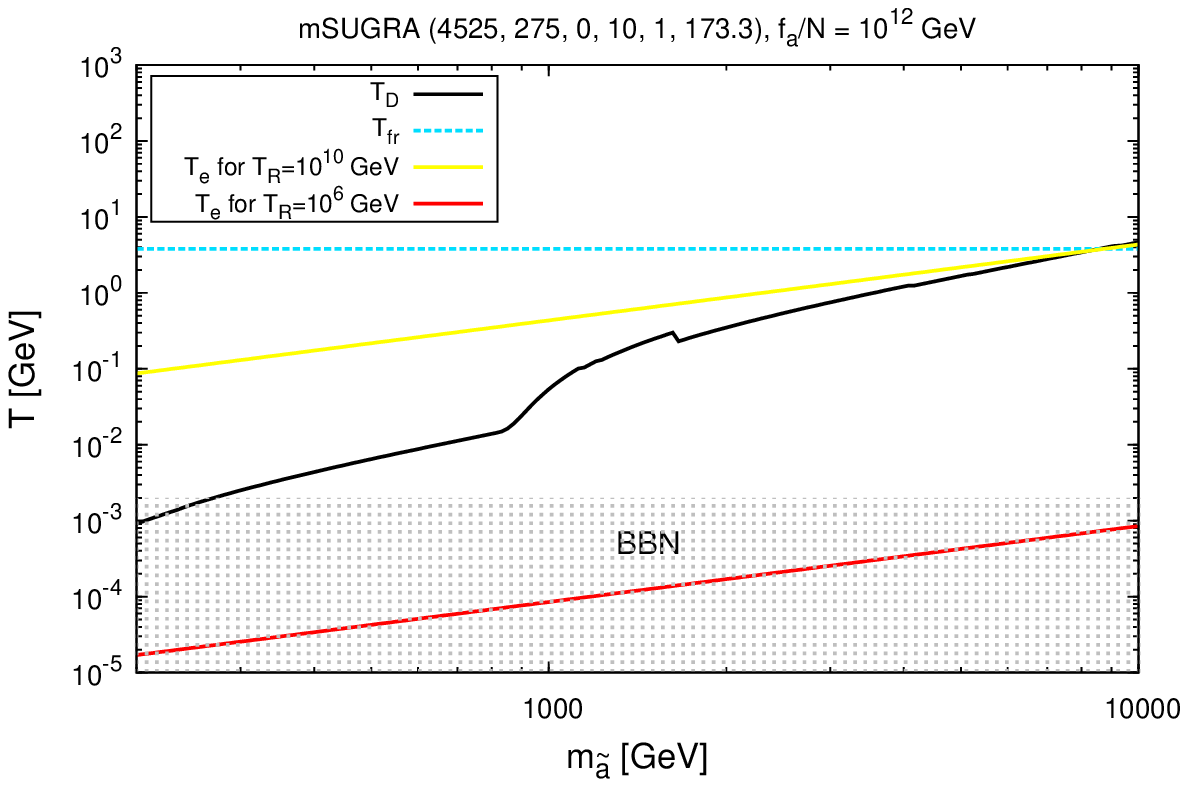}
\includegraphics[width=10cm]{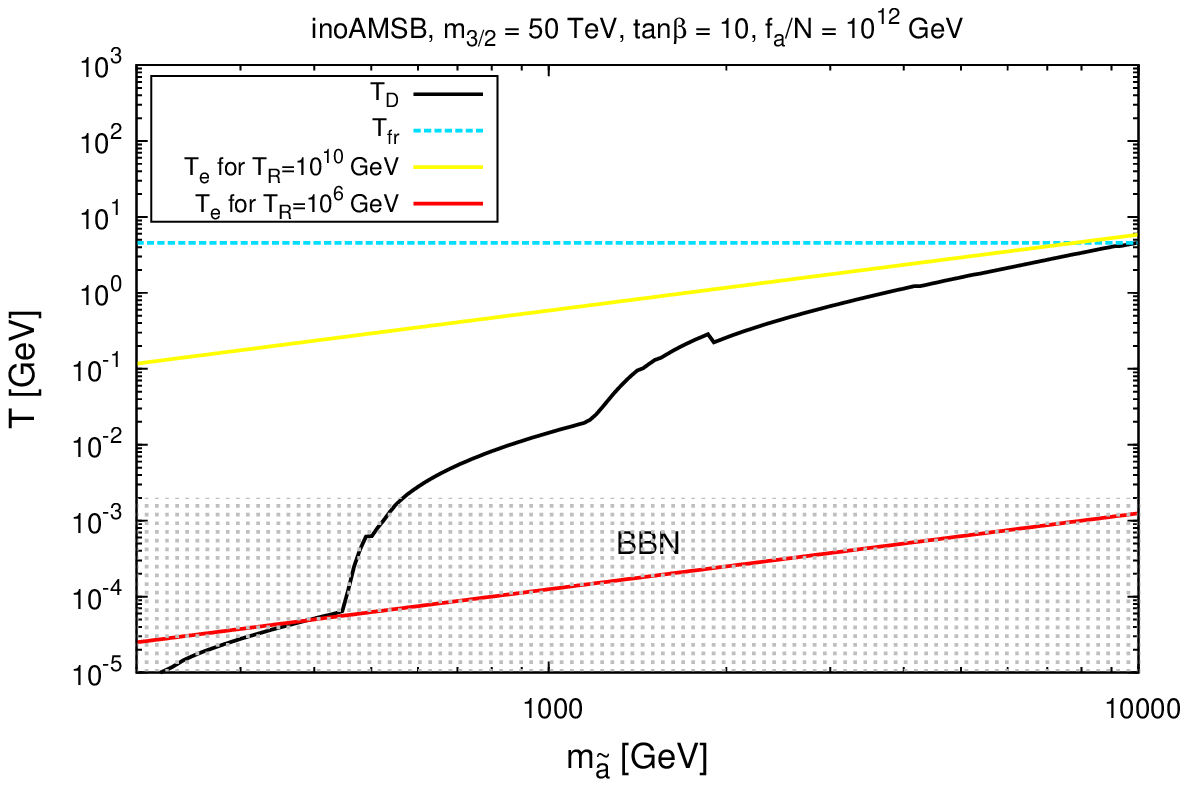}
\caption{Plot of $T_D$, $T_e$ and $T_{fr}$  versus $m_{\ta}$ in the
BM1 and BM2 benchmark models with $f_a/N=10^{12}$ GeV.
}\label{fig:TD11}}
%%%%%%%%%%%%%%%%%%%%%%%%%%%%%%%%%%%%%%%%%%%%%%%%%%%%%%%%%%%%%%%%%%%

In frame {\it b}). of Fig. \ref{fig:TD11}, we show the same temperatures, but this time for
BM2, the inoAMSB case with wino dark matter. In this case, the region with 
$m_{\ta}\alt 550$ GeV is excluded by BBN constraints since the axino width is suppressed by the fact that
only the decay to $\tz_1\gamma$ is open, where the bino component of $\tz_1$ is tiny.
When the decay $\ta\to\tz_2\gamma$ turns on around $m_{\ta}\sim 450$ GeV, 
the axino width rapidly increases, and the heavy axino scenario becomes BBN-allowed.
For $T_R=10^{10}$ GeV, axino domination occurs out to 
$m_{\ta}\sim 10$ TeV, while for higher $m_{\ta}$ values, the axino decays before
neutralino freeze-out. For the $T_R=10^6$ GeV case, axino domination only occurs
in the BBN excluded region.

%%%%%%%%%%%%%%%%%%%%%%%%%%%%%%%%%%%%%%%%%%%%%%%%%%%%%%%%%%%%%%%%%%%
%\FIGURE[t]{
%\includegraphics[width=10cm]{temp_msugra_fa12.eps}
%\includegraphics[width=10cm]{temp_ino_fa12.eps}
%\caption{Plot of $T_D$ and $T_e$  versus $m_{\ta}$ in the
%FP and inoAMSB models with $f_a/N=10^{12}$ GeV.
%}\label{fig:TD12}}
%%%%%%%%%%%%%%%%%%%%%%%%%%%%%%%%%%%%%%%%%%%%%%%%%%%%%%%%%%%%%%%%%%%

In order to see how large the increase in entropy due to axino
decay can be, we plot in Fig. \ref{fig:r} regions of $r$ ranging in value from 1
to $> 10^4$ in the $f_a/N\ vs.\ T_R$ plane for $m_{\ta}=1$ TeV and 
{\it a}). the BM1 benchmark and {\it b}). the BM2 benchmark. 
The region with $f_a/N \agt 10^{13}$ GeV is likely BBN excluded since the
axino decay temperature $T_D$ drops below $\sim 2$ MeV. From Fig. \ref{fig:r}, 
we see that when $T_R<T_{dcp}$, and axinos are produced via 
Eq. \ref{eq:axinoprod},
$r$ decreases with increasing $f_a/N$. Also, $r$ increases 
with increasing $T_R$ due to enhanced thermal production of axinos.
In contrast, when $T_R>T_{dcp}$, the axinos are produced in thermal equilibrium and the 
production rate is independent of $f_a/N$. 
In this case, the $r$ contours increase with increasing $f_a/N$, while they are
also independent of $T_R$.   

%%%%%%%%%%%%%%%%%%%%%%%%%%%%%%%%%%%%%%%%%%%%%%%%%%%%%%%%%%%%%%%%%%%
\FIGURE[t]{
\includegraphics[width=10cm]{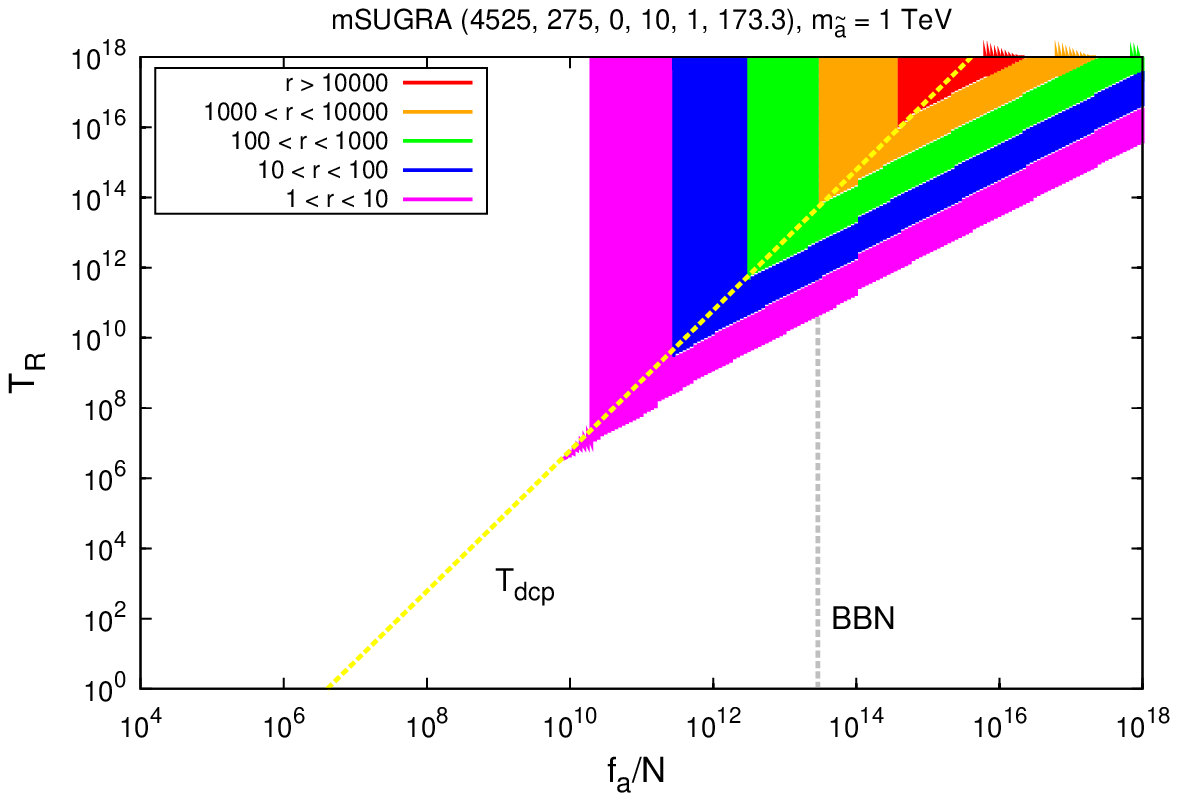}
\includegraphics[width=10cm]{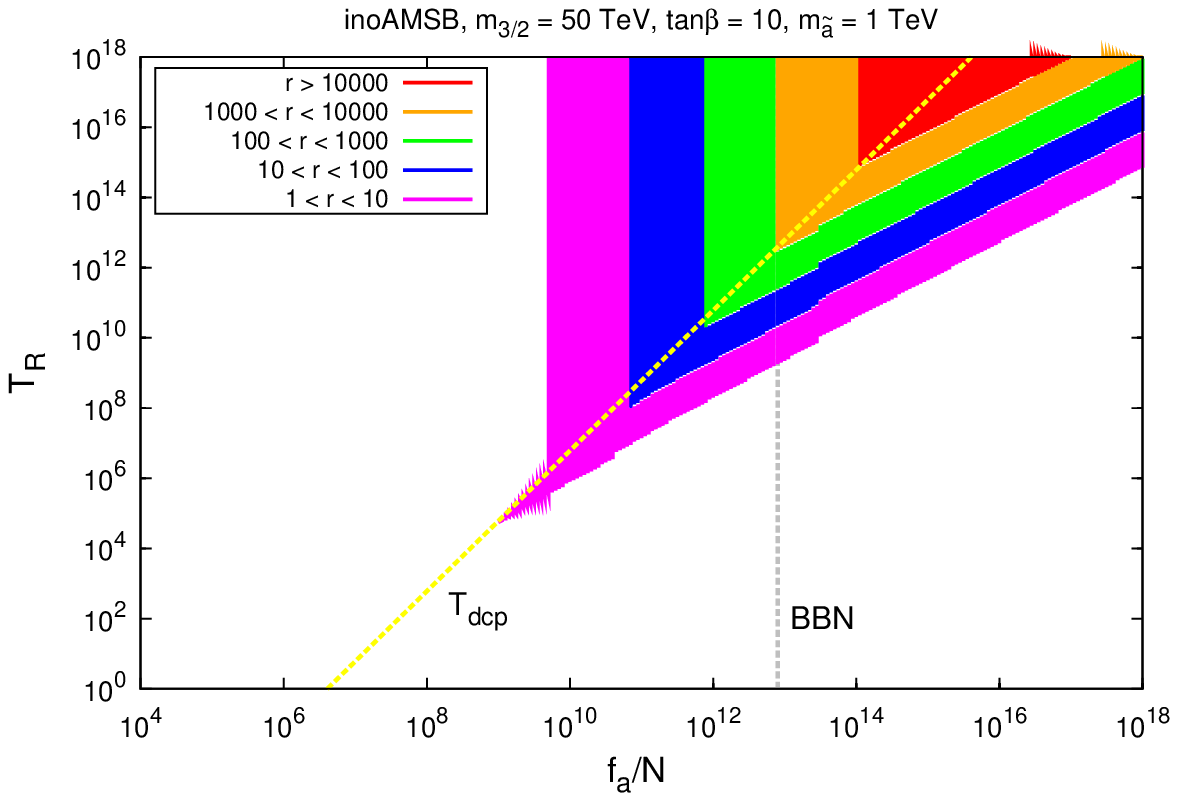}
\caption{Plot of $r$ values in the $f_a/N\ vs.\ T_R$ plane
for {\it a}) the FP model and {\it b}) the inoAMSB model,
with $m_{\ta}=1$ TeV.
}\label{fig:r}}
%%%%%%%%%%%%%%%%%%%%%%%%%%%%%%%%%%%%%%%%%%%%%%%%%%%%%%%%%%%%%%%%%%%

\section{Axion production from vacuum misalignment}
\label{sec:vacmis}

As already mentioned, the axion will contribute to the total dark matter, so we
wish to review how it is produced. We consider the scenario where the PQ 
symmetry breaks before the end of inflation, so that a nearly uniform value of 
the axion field $\theta_i\equiv a(x)/(f_a/N)$ is expected throughout the universe.
As implied by its equation of motion, the axion field stays relatively
constant until temperatures approach the QCD scale $T_{QCD}\sim 1$ GeV.
At this point, a temperature-dependent axion mass term turns on, and a 
potential is induced for the axion field. 
At temperature $T_a$ the axion field begins to oscillate,
filling the universe with low energy (cold) axions. 
The standard axion relic density (via this vacuum mis-alignment mechanism)
is derived assuming that coherent oscillations begin in a radiation-dominated 
(RD) universe ($T_a < T_D$ or $T_a > T_e$), and its final form is given 
by~\cite{vacmis,vg1}
\be
\Omega_a^{std} h^2\simeq 0.23 f(\theta_i)\theta_i^2 
\left(\frac{f_a/N}{10^{12}\ {\rm GeV}}\right)^{7/6}
\label{eq:Oh2axionstd}
\ee
where $0< \theta_i<\pi$ and $f(\theta_i)$ is the anharmonicity
factor. Visinelli and Gondolo~\cite{vg1} parametrize the latter as
$f(\theta_i)=\left[\ln\left(\frac{e}{1-\theta_i^2/\pi^2}\right)\right]^{7/6}$.
The uncertainty in $\Omega_a h^2$ from vacuum mis-alignment is estimated 
as plus-or-minus a factor of three. 

However, if the axion oscillation starts during the matter dominated (MD) or the
decaying particle dominated (DD) phase ($T_D < T_a < T_e$), the axion relic
density will no longer be given by Eq.~\ref{eq:Oh2axionstd}. The appropriate
expressions for each of these cases are given in Appendix \ref{app:AO}.

%For our purposes, we must also evaluate the axion density in a MD and
%DPD universe, and in the case where oscillations begin before
%the matter dominated phase. Expressions for each of these cases are given in Appendix \ref{app:AO}.
%We thus have
%\be
%\Omega_ah^2 = \left\{ \begin{array}{ll}
%\Omega_a^{std}h^2/r \;  \mbox{, if $T_e < T_a^{std}$} \\
%\Omega_a^{MD}h^2\;  \mbox{, if $T_{S} < T_a^{MD} < T_e$} \\
%\Omega_a^{DPD}h^2\;  \mbox{, if $T_D < T_a^{DPD} < T_{S}$} \\
%\Omega_a^{std}h^2 \;  \mbox{, if $T_a^{DPD} < T_D$}
%\end{array} \right. ,
%\label{eq:Oh2axion}
%\ee
%where $r$ is the entropy injection ratio as usual.

\section{Relic abundance of neutralinos}
\label{sec:z1}

In Ref. \cite{ckls}, formulae for the relic abundance of neutralinos
are derived which include the effects of enhancement from axino
production and decay and diminution from the re-annihilation effect
and entropy dilution.
The starting point to evaluate the neutralino abundance is the Boltzmann
equation
\be
\frac{dn_{\tz_1}}{dt}+3Hn_{\tz_1}=-\langle\sigma v\rangle n_{\tz_1}^2 ,
\label{eq:boltz}
\ee
where $n_{\tz_1}$ is the neutralino number density, $H(t)$ is the Hubble constant at time $t$
($H^2=\rho (t)/3M_{P}^2$), 
$\sigma$ is the neutralino annihilation cross section, $v$ is the 
$\tz_1-\tz_1$ relative velocity and $\langle\cdots\rangle$ denotes thermal averaging.
At very early times and high temperatures, $n_{\tz_1}$ is given by the thermal
equilibrium abundance, but as the universe expands and cools, at a temperature $T_{fr}$ 
the expansion rate $H$ outstrips the annihilation rate 
$n_{\tz_1}\langle\sigma v\rangle$, and a relic abundance of neutralinos freezes out. 
Thus, we define here the freeze-out temperature by the value of $T$ at which
\be
\langle\sigma v \rangle n_{\tz_1}\simeq H(T_{fr}) .
\ee
From this condition we can compute the yield variable $Y=n_{\tz_1}/s$:
\be
Y_{\tz_1}^{fr}(T_{fr}) =k_{fr} \times \frac{n_{\tz_1}}{s}=\frac{H(T_{fr})}{\langle\sigma v\rangle s} .
\ee
where $k_{fr} = 1 (3/2)$ if the freeze-out happens in the radiation (matter) dominated phase.
If we assume a nearly constant value of
$\langle\sigma v\rangle$ (which is appropriate for a higgsino- or wino-like $\tz_1$
which dominantly annihilates via $s$-wave reactions), and freeze-out in a radiation-dominated
universe with $H^2=\rho_{rad}/3M_P^2$ and $\rho_{rad}=\pi^2 g_*T^4/30$, 
then one obtains
\be
Y_{\tz_1}^{fr}=\frac{\left(90/\pi^2 g_*(T_{fr})\right)^{1/2}}{4 \langle\sigma v\rangle M_PT_{fr}} .
\label{eq:yield}
\ee
The usual cosmological assumption is that the yield $Y_{\tz_1}^{fr}$ is conserved from $T=T_{fr}$ to
$T_0$, where $T_0$ is the present day temperature of radiation $T_0= 2.725^\circ$K. In this case
the thermal neutralino relic density is simply
\be
\Omega_{\tz_1}^{fr} h^2 = \frac{2 \pi^2}{45} \frac{g_*(T_0) T_0^3}{\rho_c/h^2} m_{\tz_1} Y_{\tz_1}^{fr}
\ee
However, just as in the axion case, we must re-evaluate $Y_{\tz_1}^{fr}$
at $T=T_D$, and also the freeze-out temperature, in the cases of a MD or a DD universe.
The various yield expressions are contained in Appendix \ref{app:z1}.
%Here, we merely list the final expressions for the neutralino abundane 
%depending on whether the universe is RD, MD or DD:
%\be
%\Omega_{\tz_1} h^2 = \left\{ \begin{array}{ll}
%\frac{1}{r} 8.5\times10^{-11} m_{\tz_1} \frac{1}{\sqrt{g_*(T_{fr}^{std})}} \frac{1}{T_{fr}^{std}}\frac{1}{\langle \sigma v\rangle} \;  \mbox{, if $T_e < T_{fr}^{std}$} \\
%\frac{1}{r}\frac{3}{2} \times 8.5\times10^{-11} m_{\tz_1} \frac{1}{\sqrt{g_*(T_{fr}^{MD})}} 
%\frac{\sqrt{T_e}}{(T_{fr}^{MD})^{3/2}} \frac{1}{\langle \sigma v\rangle}\;  
%\mbox{, if $T_{S} < T_{fr}^{MD} < T_e$} \ \ (MD)\\
%\frac{3}{2} \times 8.5\times10^{-11} m_{\tz_1} \frac{\sqrt{g_*(T_D)}}{g_*(T_{fr}^{DPD})} 
%\frac{T_D^3}{(T_{fr}^{DPD})^{4}} \frac{1}{\langle \sigma v\rangle}\;  
%\mbox{, if $T_D < T_{fr}^{DPD} < T_{S}$} \ \ (DPD)\\
%8.5\times10^{-11} m_{\tz_1} \frac{1}{\sqrt{g_*(T_{fr}^{std})}} \frac{1}{T_{fr}^{std}}
%\frac{1}{\langle \sigma v\rangle}\;  \mbox{, if $T_{fr}^{DPD} < T_D$\ \ (RD)}
%\end{array} \right.
%\ee
%where $\langle \sigma v\rangle$ must be evaluated at the appropriate $T_{fr}$ expression, as
%given in Appendix \ref{app:z1}.

\subsection{Neutralino re-annihilation at $T=T_D$}
\label{ssec:re-ann}

In the mixed $a\tz_1$ DM scenario, neutralinos are produced via axino decay at 
temperature $T=T_D$, as well as via thermal freeze-out at $T=T_{fr}$. The needed condition for 
re-annihilation at $T\sim T_D$ is that the annihilation rate exceeds the expansion rate at $T=T_D$:
\be
\langle\sigma v\rangle \frac{n_{\tz_1}(T_D)}{s} > \frac{H(T_D)}{s}
\ee
where $n_{\tz_1}(T_D)$ is the total neutralino number density due to
thermal (freeze-out) and non-thermal (axino decays) production. We can
rewrite the above condition as:
\be
Y_{\tz_1}|_{T=T_D}=\left(Y_{\tz_1}^{\ta}+Y_{\tz_1}^{fr} (T_D)\right)  > 
\frac{\left( 90/\pi^2g_*(T_D)\right)^{1/2}}{4\langle\sigma v\rangle M_P T_D}  ,
\label{eq:13}
\ee
where for $r>1$ we have $Y_{\tz_1}^{\ta} = Y_{\ta}/r$ and $Y_{\tz_1}^{fr}(T_D)$ is given by Eq. \ref{eq:Yrgt1}. 
If this condition is satisfied, then additional neutralino annihilation will occur at $T\sim T_D$.

The relevant regions are shown in the $T_R \ vs. \ m_{\ta}$ plane for
$f_a/N=10^{12}$ GeV in Fig. \ref{fig:tr_ma_msugra} for BM1 and in
Fig. \ref{fig:tr_ma_ino} for BM2. The magenta, blue and green regions show ranges
of different entropy dilution factors $r$, while the brown-shaded region is where
there are not enough neutralinos produced through axino decays to cause 
re-annihilation. The region to the left of the BBN line is excluded, since
$T_D<2$ MeV, and thus we see that for nearly all of the allowed parameter space,
neutralino re-annihilation effects need to be included in our calculations of the
neutralino relic density.  We also show the (vertical, dotted) line where 
$T_D =T_{fr}$. The region to the right of this line is where $T_D >T_{fr}$ and the 
axino decay products are expected to thermalize before neutralino freeze-out .
%
%%%%%%%%%%%%%%%%%%%%%%%%%%%%%%%%%%%%%%%%%%%%%%%%%%%%%%%%%%%%%%%%%%%
\FIGURE[t]{
\includegraphics[width=10cm]{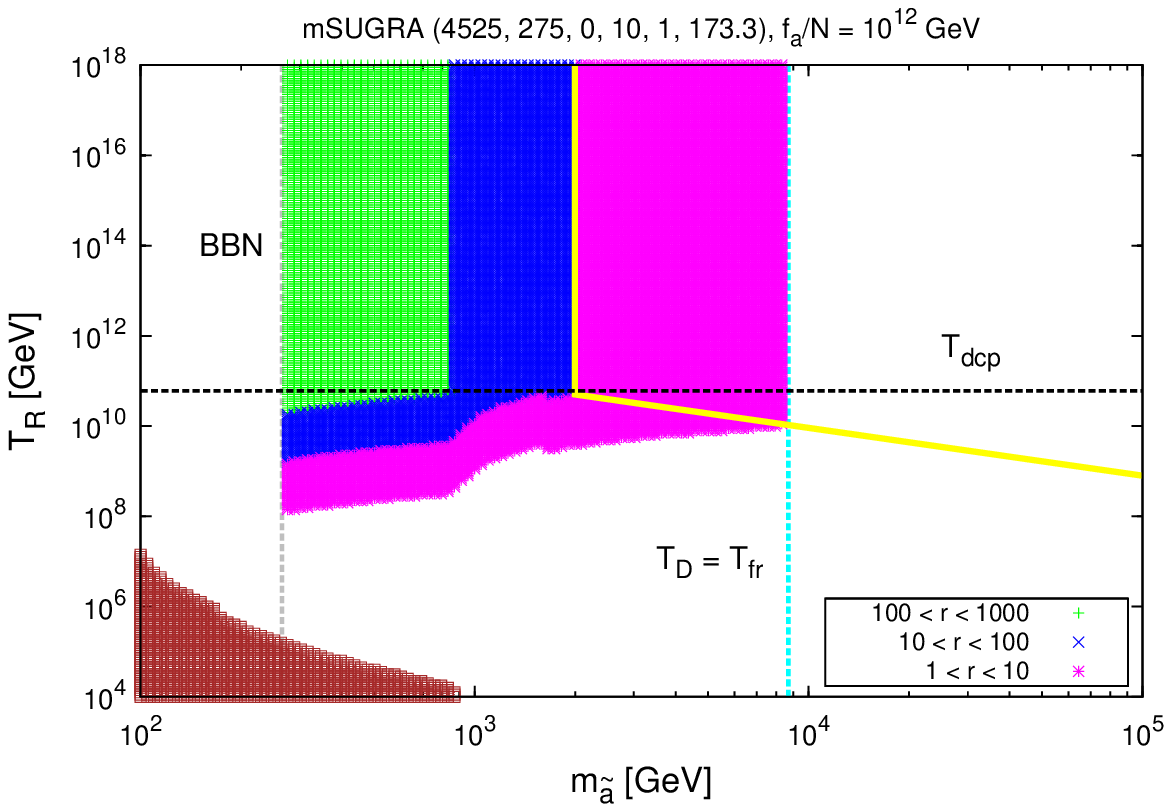}
\caption{Plot of regions of $m_{\ta}\ vs.\ T_R$
plane where $T_{fr}<T_e$
(right of yellow contour) and
where {\it no} additional neutralino annihilation occurs (brown), 
in the BM1 (HB/FP) benchmark model.
We also show regions of entropy generation $r$.
}\label{fig:tr_ma_msugra}}
%%%%%%%%%%%%%%%%%%%%%%%%%%%%%%%%%%%%%%%%%%%%%%%%%%%%%%%%%%%%%%%%%%%
%
%%%%%%%%%%%%%%%%%%%%%%%%%%%%%%%%%%%%%%%%%%%%%%%%%%%%%%%%%%%%%%%%%%%
\FIGURE[t]{
\includegraphics[width=10cm]{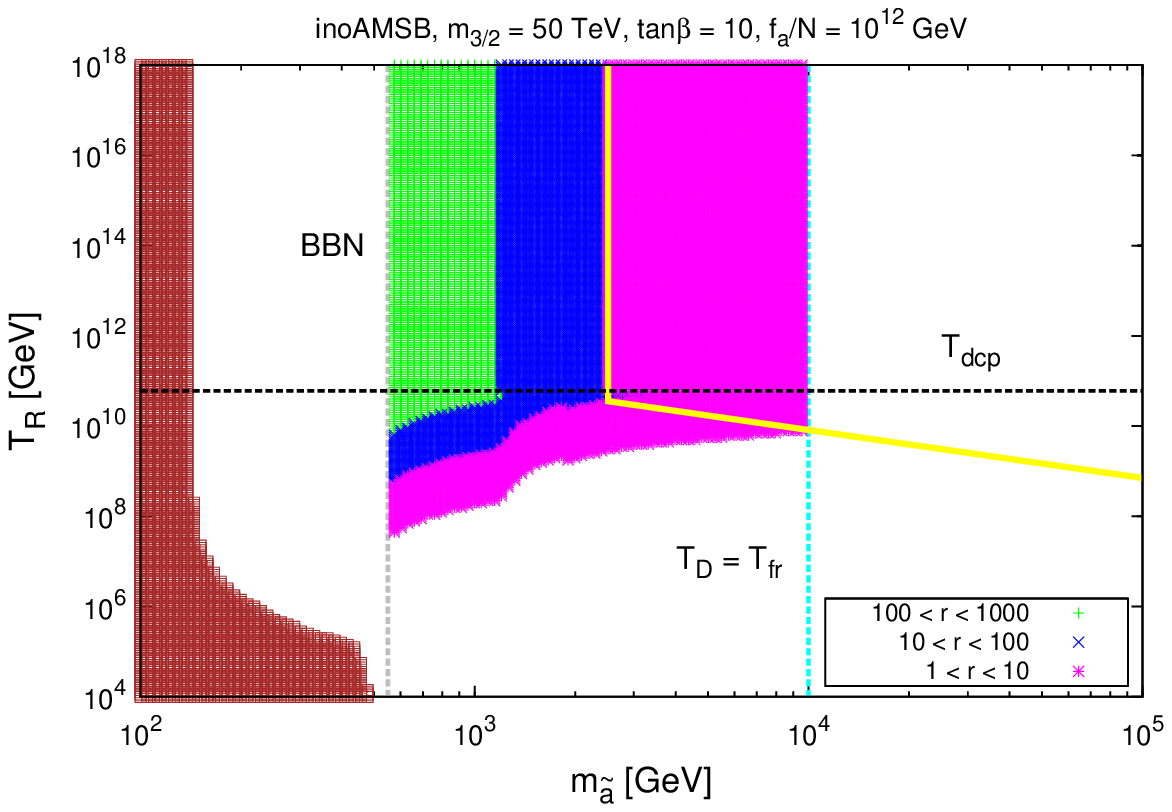}
\caption{Plot of regions of $m_{\ta}\ vs.\ T_R$
plane where $T_{fr}<T_e$
(right of yellow contour) and
where {\it no} additional neutralino annihilation occurs (brown), 
in the BM2 (inoAMSB) benchmark model.
We also show regions of entropy generation $r$.
}\label{fig:tr_ma_ino}}
%%%%%%%%%%%%%%%%%%%%%%%%%%%%%%%%%%%%%%%%%%%%%%%%%%%%%%%%%%%%%%%%%%%

A simple expression for the diminution of neutralinos from re-annihilation has been
worked out in Ref. \cite{ckls} in the sudden-decay approximation. 
Recasting the Boltzmann equation \ref{eq:boltz} in terms of the yield variable
\be
\frac{dY_{\tz_1}}{dt}=-\langle\sigma v\rangle Y_{\tz_1}^2s
\ee
and integrating from time $t=t_D$ to $t$ gives
\be
Y_{\tz_1}^{-1}(T)=Y_{\tz_1}^{-1}(T_D)-\langle\sigma v\rangle\left(\frac{s}{H}-\frac{s(T_D)}{H(T_D)}\right)
\simeq Y_{\tz_1}^{-1}(T_D)+\frac{\langle\sigma v\rangle s(T_D)}{H(T_D)} .
\label{eq:reann}
\ee
where $Y_{\tz_1}(T_D)=Y_{\tz_1}^{fr}+Y_{\tz_1}^{\ta}$, and for $T_R\agt 10^4$ GeV, the
decay contribution can be huge, so that $Y_{\tz_1}^{-1}(T_D)$ is small. Then Eq. \ref{eq:reann} is dominated
by the second term and
\be
Y_{\tz_1}(T)\simeq H(T_D)/\langle\sigma v\rangle s(T_D)\simeq Y_{\tz_1}^{fr} (T_D)
\ee
Since the latter term
is evaluated at $T=T_D$, it is much larger than $Y_{\tz_1}^{fr}(T_{fr})$, and in fact the neutralino
abundance turns out to be nearly the freeze-out abundance but evaluated at the much lower temperature $T_D$.
%This is what is meant by $T_D$ effectively acting as a second re-heat temperature.

To exhibit the effect graphically, in Fig. \ref{fig:Y} we plot the neutralino yield $Y$ versus
$T_R$ for {\it a}). the BM1 benchmark and {\it b}). the BM2 benchmark.
In both cases, we take $f_a/N=10^{12}$ GeV and $m_{\ta}=1$ TeV. The blue curve 
represents the sum of the thermal neutralino yield from freeze-out plus the neutralino yield
from axino production and decay, where this term dominates for all but the lowest values of
$T_R$ in the plot. It increases with $T_R$ as the thermal yield of axinos, Eq. \ref{eq:axinoprod},
increases linearly with $T_R$. For $T_R$ high enough that axinos can dominate the universe, 
the neutralino abundance is diluted by entropy production by the same factor as axinos are produced, 
and thus the curve becomes flat. The green curve is the second term of Eq. \ref{eq:reann} and corresponds to
the thermal neutralino yield, Eq. \ref{eq:yield}, but evaluated at $T_D$ instead of $T_{fr}$. 
The red curve denotes the final
neutralino yield, {\it i.e.} that given by Eq. \ref{eq:reann}.
From Fig. \ref{fig:Y} we see that, for $T_R \simeq 10^4$~GeV, the final neutralino yield ($Y_{\tz_1}(T)$)
is close to the naive sum of thermal and non-thermal production ($Y_{\tz_1}(T_D)$) with a small suppression
due to the re-annihilation effect. As $T_R$ increases, $Y_{\tz_1}^{\ta}$ increases and so does $Y_{\tz_1}(T)$, but
at a smaller rate, since the re-annihilation becomes more and more efficient. Once $T_R \simeq 10^5$~GeV,
the annihilation term is efficient enough to re-annihilate all neutralinos down to their equilibrium value ($Y_{\tz_1}^{fr}(T_D)$).
Thus, for $T_R \gtrsim 10^5$~GeV, the neutralinos injected from axino decays can effectively thermalize to 
$Y_{\tz_1}^{fr}(T_D)$ and the final neutralino yield becomes independent of the initial axino abundance.
%For high $T_R$, it tracks the green curve, 
%denoting that the final neutralino yield is essentially the thermal yield evaluated at the second
%reheat temperature $T_D$. At low $T_R$, the total yield is lower than $Y_{\tz_1}^{fr}(T_D)$ since in this
%case the thermal axino abundance is also low, and the two terms in Eq. \ref{eq:reann} are comparable.
Since $T_D$ is usually far below $T_{fr}$, Eq. \ref{eq:yield} tells us that the neutralino
yield $Y_{\tz_1}(T) \simeq Y_{\tz_1}^{fr}(T_D)$
will likely far exceed the thermal yield evaluated at $T_{fr}$, $Y_{\tz_1}^{fr}(T_{fr})$.
%
%%%%%%%%%%%%%%%%%%%%%%%%%%%%%%%%%%%%%%%%%%%%%%%%%%%%%%%%%%%%%%%%%%%
\FIGURE[t]{
\includegraphics[width=10cm]{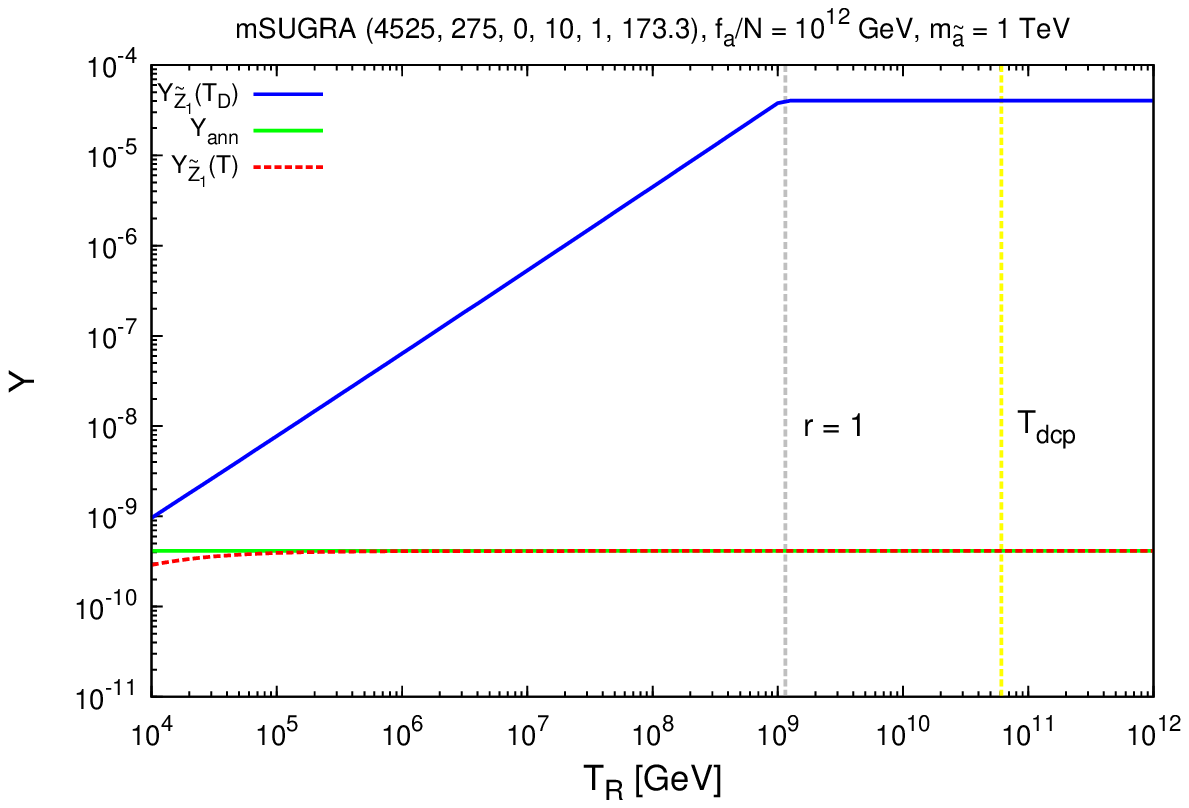}
\includegraphics[width=10cm]{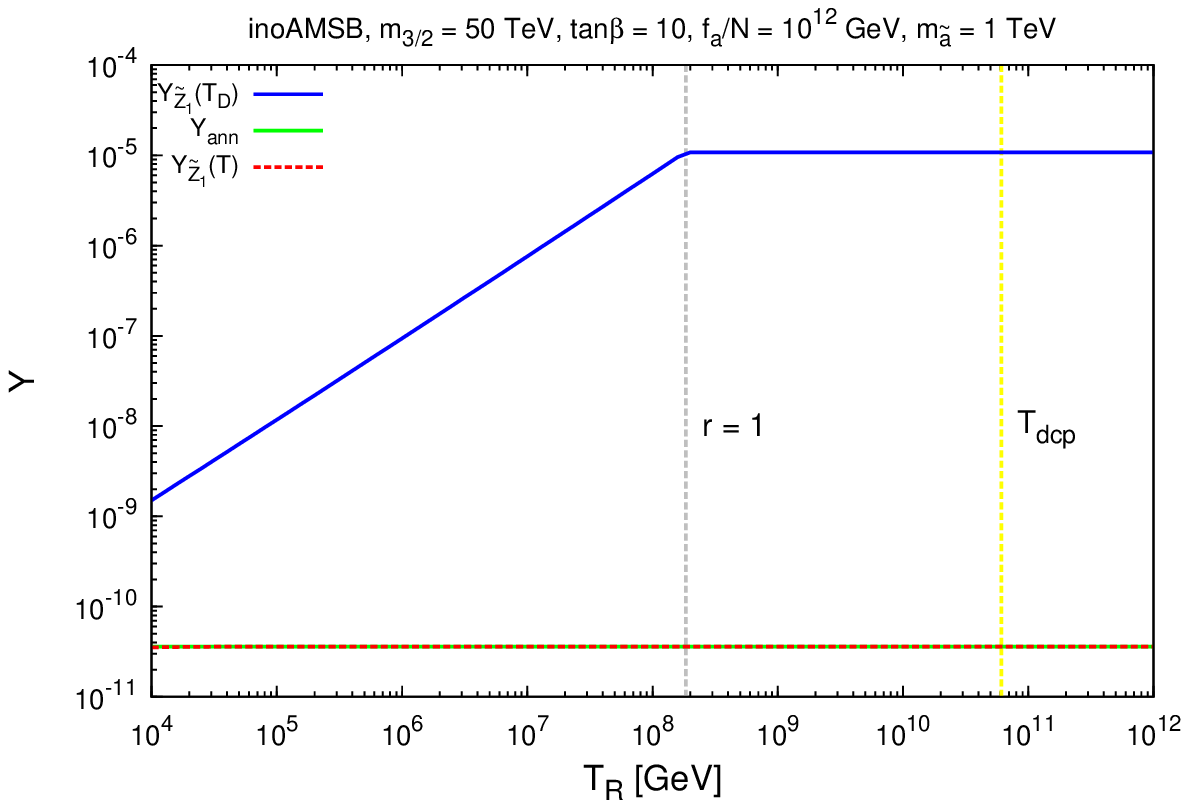}
\caption{Plot of yield $Y$ versus $T_R$ for $f_a/N=10^{12}$ GeV and $m_{\ta}=1$ TeV
for {\it a}) the FP model and {\it b}) the inoAMSB model.
}\label{fig:Y}}
%%%%%%%%%%%%%%%%%%%%%%%%%%%%%%%%%%%%%%%%%%%%%%%%%%%%%%%%%%%%%%%%%%%

%
%%%%%%%%%%%%%%%%%%%%%%%%%%%%%%%%%%%%%%%%%%%%%%%%%%%%%%%%%%%%%%%%%%%
\FIGURE[t]{
\includegraphics[width=10cm]{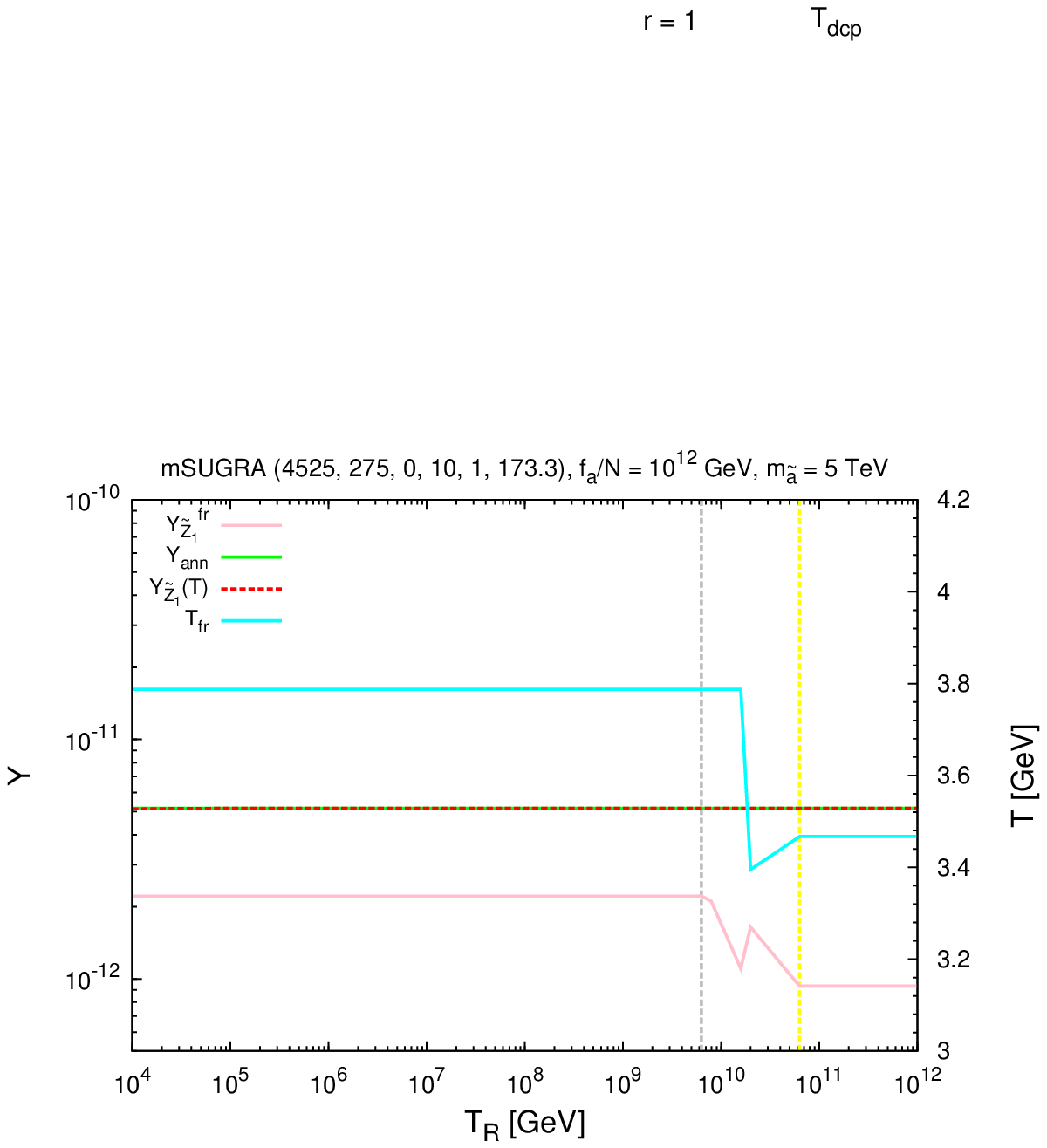}
\caption{Plot of yield $Y$ versus $T_R$ for $f_a/N=10^{12}$ GeV and $m_{\ta}=5$ TeV
for the BM1 (HB/FP) model.  The dashed gray line shows where $r=1$ and the yellow
line shows where $T_R = T_{dcp}$.
}\label{fig:Ym5}}
%%%%%%%%%%%%%%%%%%%%%%%%%%%%%%%%%%%%%%%%%%%%%%%%%%%%%%%%%%%%%%%%%%%

For benchmarks BM1 and BM2, the region to the right of the yellow contour and to
the left of the vertical blue line denotes where, in Figs. 
\ref{fig:tr_ma_msugra} and \ref{fig:tr_ma_ino}, we have $T_D < T_{fr} < T_e$.
In this case, the thermal neutralino abundance must be calculated in a MD or DD universe.
To illustrate, we show the various neutralino yields in Fig. \ref{fig:Ym5} for benchmark BM1
with $m_{\ta}=5$ TeV. The pink curve denotes the thermal neutralino abundance, 
which is constant versus $T_R$ until $r>1$, where the abundance becomes diluted by $1/r$
and begins to decrease. At $T_R\sim 2\times 10^{10}$ GeV, $T_e$ begins to exceed $T_{fr}$,
the neutralino freeze-out happens in a MD universe, and suffers an increase.
The corresponding freeze-out temperature is denoted by the azure colored curve, and the right-hand
$y$-axis: we see that $T_{fr}$ drops from $\sim 3.8$ GeV to $\sim 3.5$ GeV as we move
from a RD to a MD universe. From Fig. \ref{fig:Ym5} we see that the change in $Y_{\tz_1}^{fr}$ is
inconsequential to the final neutralino yield ($Y_{\tz_1}(T)$), since this is
%The ultimate effect in this case is inconsequential since the
%thermal abundance is dwarfed by neutralino production from axino decay (dark blue curve).
%The ultimate abundance is then 
determined by the re-annihilation term (green curve), and is
insensitive to the thermal neutralino abundance, at least in this case.

%If $T_e<T_{fr}$, then the thermal yield is given by the usual expression \ref{eq:yield}.
%The yield for neutralinos from axino production and decay is given by 
%\be
%Y_{\tz_1}^{\ta}=Y_{\ta}/(S_f/S_0) ,
%\label{eq:yieldent}
%\ee
%{\it i.e.} the neutralinos inherit the axino number density after decay, but the yield is
%diluted by entropy production from axino decay.

%\subsection{Relic Density}

%If there is no re-annihilation of neutralinos at $T=T_D$, the neutralino relic
%density is simply:
%\be
%\Omega_{\tz_1} h^2 = \frac{2 \pi^2}{45} \frac{g_*(T_0) T_0^3}{\rho_c/h^2} m_{\tz_1} 
%Y_{\tz_1}(T_D) 
%\ee
%Hence:
%\be
%\Omega_{\tz_1} h^2 = \left\{ \begin{array}{ll}
%\frac{1}{r} 8.5\times10^{-11} m_{\tz} \frac{1}{\sqrt{g_*(T_{fr}^{std})}} \frac{1}{T_{fr}^{std}}\frac{1}{\langle \sigma v\rangle} \;  \mbox{, if $T_e < T_{fr}^{std}$} \\
%\frac{1}{r}\frac{3}{2} \times 8.5\times10^{-11} m_{\tz} \frac{1}{\sqrt{g_*(T_{fr}^{MD})}} 
%\frac{\sqrt{T_e}}{(T_{fr}^{MD})^{3/2}} \frac{1}{\langle \sigma v\rangle}\;  
%\mbox{, if $T_{S} < T_{fr}^{MD} < T_e$} \\
%\frac{3}{2} \times 8.5\times10^{-11} m_{\tz} \frac{\sqrt{g_*(T_D)}}{g_*(T_{fr}^{DPD})} 
%\frac{T_D^3}{(T_{fr}^{DPD})^{4}} \frac{1}{\langle \sigma v\rangle}\;  
%\mbox{, if $T_D < T_{fr}^{DPD} < T_{S}$} \\
%8.5\times10^{-11} m_{\tz} \frac{1}{\sqrt{g_*(T_{fr}^{std})}} \frac{1}{T_{fr}^{std}}
%\frac{1}{\langle \sigma v\rangle}\;  \mbox{, if $T_{fr}^{DPD} < T_D$}
%\end{array} \right.
%\ee
%where $\langle \sigma v\rangle$ must be evaluated at the appropriate $T_{fr}$ as given in 
%Appendix \ref{app:z1}.

\section{Algorithm for DM abundance in $a\tz_1$ DM scenario}
\label{sec:algor}

In this section, we list our algorithm for evaluating the relic density of mixed
$a\tz_1$ CDM in the PQMSSM. We proceed as follows.
First, extract an effective value of $\langle\sigma v\rangle$ by matching Eq. \ref{eq:yield}
onto the $\Omega_{\tz_1}h^2$ result from the IsaReD subroutine of Isajet (the effective
$\langle\sigma v\rangle$ value is shown in Table \ref{tab:bm}).
Then:
\bi
\item If $T_D< T_{BBN}$ (where $T_{BBN}$ is taken as 2 MeV), regard as BBN excluded\cite{jedamzik}.
\item If $T_D>T_{fr}$, then $\Omega_{\tz_1}h^2 = \Omega_{\tz_1}^{std} h^2$, given by the IsaReD output.
\item If $T_D<T_{fr}$ and $T_D>T_{BBN}$, then
\bi
\item If $r>1$: (Case of axino domination with $T_e>T_D$)
\bi
\item If $\tz_1$ re-annihilate (Eq. \ref{eq:13} satisfied),
then calculate $Y_{\tz_1}$ using Eq. \ref{eq:reann} with $Y_{\tz_1}(T_D)=Y_{\tz_1}^{fr}+Y_{\tz_1}^{\ta}$,
where $Y_{\tz_1}^{fr}$ is given by Eq. \ref{eq:Yrgt1} and $Y_{\tz_1}^{\ta}= Y_{\ta}/r$.

\item If $\tz_1$ does not re-annihilate (Eq. \ref{eq:13} not satisfied), then
$Y_{\tz_1}=Y_{\tz_1}^{fr}+Y_{\tz_1}^{\ta}$,
where $Y_{\tz_1}^{fr}$ is given by Eq. \ref{eq:Yrgt1} and $Y_{\tz_1}^{\ta}= Y_{\ta}/r$. 
\ei
\item If $r<1$: (axino non-domination)
\bi
\item If neutralinos re-annihilate (Eq. \ref{eq:13} holds), then yield
given by Eq. \ref{eq:reann} with $Y_{\tz_1}^{fr}$ given by Eq. \ref{eq:yield} and
$Y_{\tz_1}^{\ta}=Y_{\ta}$.
\item If neutralinos do not re-annihilate (Eq. \ref{eq:13} does not hold), then yield
given by $Y_{\tz_1}^{fr}+Y_{\tz_1}^{\ta}$ with $Y_{\tz_1}^{fr}$ 
given by Eq. \ref{eq:yield} and $Y_{\tz_1}^{\ta}= Y_{\ta}$.
\ei
\ei
\item Now add in axion contribution to relic abundance Eq. \ref{eq:Oh2axion}
\item Final dark matter abundance is given by: 
$\Omega_{a\tz_1}h^2=\Omega_{\tz_1}h^2+\Omega_a h^2$.
\ei

\section{Numerical results}
\label{sec:results}

In this section, we present results for the relic abundance of mixed
$a\tz_1$ CDM for benchmarks BM1 and BM2 in the PQMSSM. 
Our first results are shown in Fig. \ref{fig:Oh2tot_max},
where we plot the neutralino abundance, $\Omega_{\tz_1}h^2$, the axion 
abundance, $\Omega_a h^2$, and their combination, $\Omega_{a\tz_1}h^2$,
versus $m_{\ta}$ for $f_a/N=10^{12}$ GeV, with $T_R=10^{10}$ GeV.
We take the initial axion field value $\theta_i =0.498$ for the
BM1 benchmark case shown in frame {\it a}). and we take $\theta_i=0.675$ for 
the BM2 benchmark in frame {\it b}). These values tune the total dark matter abundance to
the WMAP value for the case where $T_D>T_{fr}$. 
From frame {\it a})., we find the region to the left of the dashed gray
line is excluded by BBN constraints on late decaying neutral
particles, since $T_D<2$ MeV. For $m_{\ta}$ values just beyond the 
BBN constraint, the naive expectation is that the neutralino abundance is determined
by the thermal axino production rate, and indeed Fig. \ref{fig:Oh2_fa} suggests the 
abundance $\Omega_{\tz_1}^{\ta}h^2\sim 10^7$. Instead, the actual abundance is several
orders of magnitude below this, but still far above the measured DM value.
In this case, the large thermal axino production rate is followed by decays to
neutralinos at $T=T_D$.  As relic neutralinos fill the universe they proceed to 
annihilate, so that their final abundance is determined by Eq. \ref{eq:reann}. 
Since in this region $\Omega_{\tz_1}h^2\sim 1/T_D$, and since 
$T_D\sim m_{\ta}^{3/2}$, we find the neutralino abundance decreasing
with increasing $m_{\ta}$. The kink in frame {\it a}). at $m_{\ta}\sim 900$ GeV occurs 
due to turn-on of the $\ta\to\tg g$ decay mode, which increases $\Gamma_{\ta}$, thus decreasing
$\Omega_{\tz_1}h^2$ even further. The jog at $m_{\ta}\sim 1.6$ TeV is caused by
a change in the $g_*$ value due to the addition of quark and gluon degrees of freedom 
after the QCD phase transition.
While $\Omega_{\tz_1}h^2$ is decreasing with increasing $m_{\ta}$, it 
reaches $0.11$ at $m_{\ta}\sim 6$ TeV and continues dropping until $T_D$ exceeds $T_{fr}$.
At this point, the thermal $\tz_1$ abundance assumes its traditional value
of $\Omega_{\tz_1}^{std} h^2\sim 0.05$ since now axinos decay before freeze-out. 
For $m_{\ta}\agt 8.5$ TeV, the CDM is a nearly equal mix of axions and neutralinos.
While $T_D>1$ GeV (for high $m_{\ta}$), the axion abundance assumes the form as given in
Eq. \ref{eq:Oh2axionstd}. However, if considering lower values of $m_{\ta}$, 
then $r>1$ and $T_D<1$ GeV, so that the axion abundance is diluted by entropy 
production from axino decay. As $m_{\ta}$ increases, the axion dilution becomes 
lessened until $T_D$ exceeds $\sim 1$ GeV, after which the axion abundance
remains fixed.

In frame {\it b})., for the inoAMSB model, we again see that the neutralino
abundance is large at small $m_{\ta}$: $\Omega_{\tz_1}h^2\sim 10$ for $m_{\ta}\simeq 600$ GeV. 
It decreases steadily with $m_{\ta}$ due to the increasing $T_D$. 
The curve does reach a point where the neutralino abundance matches the WMAP7 measured value, 
at $m_{\ta}\sim 1.5$ TeV. 
In fact, the neutralino production via axino decay and subsequent re-annihilation effect offers
an elegant means to enhance the WIMP relic density in SUSY models with higgsino or wino-like WIMPs.
This mechanism offers an alternative\cite{shibi} to the neutralino abundance enhancement via moduli 
decays as proposed by Morio and Randall\cite{moroi} for AMSB, 
and as proposed by Kane et al. to explain the Pamela/Fermi cosmic ray anomalies\cite{kane}.
By the time $T_D$ exceeds $T_{fr}$, the axion
abundance has assumed the value given by Eq. \ref{eq:Oh2axionstd}, and
$\Omega_{\tz_1}h^2\sim 10^{-3}$, as given in Table \ref{tab:bm}.
%
%%%%%%%%%%%%%%%%%%%%%%%%%%%%%%%%%%%%%%%%%%%%%%%%%%%%%%%%%%%%%%%%%%%
\FIGURE[t]{
\includegraphics[width=10cm]{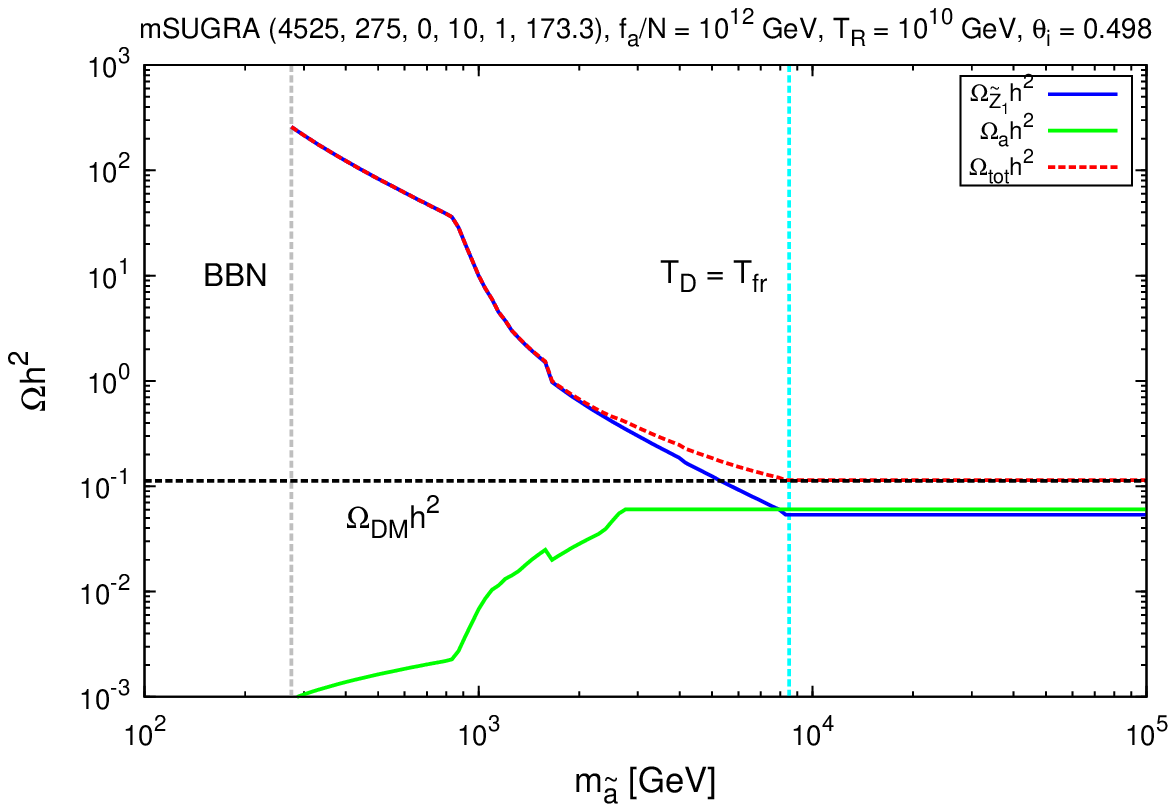}
\includegraphics[width=10cm]{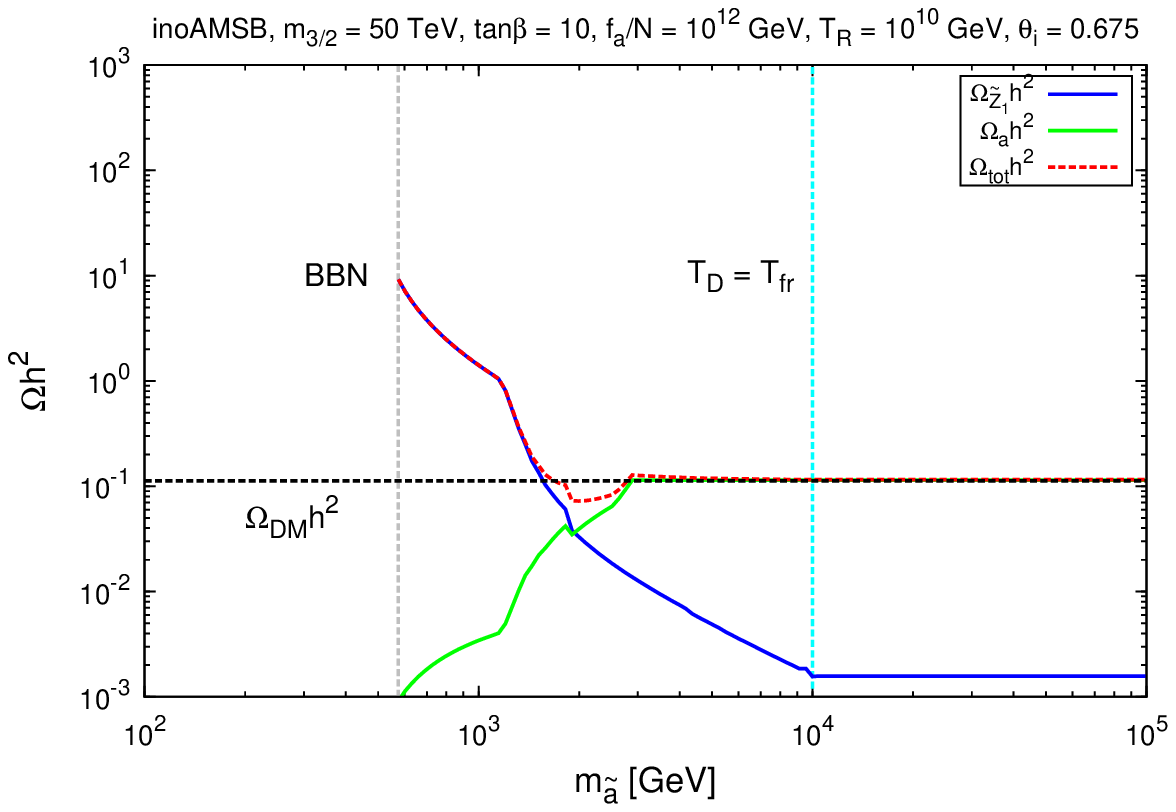}
\caption{Plot of neutralino and axion relic densities  $\Omega h^2$ versus $m_{\ta}$ for
$f_a/N=10^{12}$ GeV and $T_R=10^{10}$ GeV
for {\it a}) the HB/FP model and {\it b}) the inoAMSB model.
}\label{fig:Oh2tot_max}}
%%%%%%%%%%%%%%%%%%%%%%%%%%%%%%%%%%%%%%%%%%%%%%%%%%%%%%%%%%%%%%%%%%%

In Fig. \ref{fig:Oh2tot_fa}, we show the mixed $a\tz_1$ abundance versus $f_a/N$
for fixed $m_{\ta}=1$ TeV and $T_R=10^{10}$ GeV for {\it a}). BM1 and {\it b}). the
BM2 benchmarks. In the HB/FP case of frame {\it a})., we see that for low
$f_a/N$, the axino width $\Gamma_{\ta}$ is large, and $T_D$ exceeds $T_{fr}$, so 
that axinos decay before freeze-out and the neutralino relic density assumes
its standard value of $\Omega_{\tz_1}^{std} h^2 = 0.05$ in this case. Meanwhile, 
the axion density is extremely small due to the low value of $f_a/N$. As
$f_a/N$ increases, the axion abundance naturally increases, while the neutralino
abundance remains constant until around $f_a/N\sim 10^{10}$ GeV, where
$T_D$ falls below $T_{fr}$. For higher $f_a/N$ values, $T_D$ continues to fall and since 
$\Omega_{\tz_1}h^2\sim T_D^{-1}$, the neutralino abundance steadily increases. 
At the point where $r=1$ is reached ($f_a/N\sim 2\times 10^{10}$ GeV), 
entropy injection from axino decay causes a small decline in the
otherwise steadily increasing axion abundance.
The general behavior in frame {\it b}). for the inoAMSB model is similar to 
that of frame {\it a}). Nowhere in these two plots does the axion abundance exceed
the neutralino abundance. This is merely a reflection of our choice of $m_{\ta}=1$ TeV and $\theta_i$;
for higher values of $m_{\ta}$, the value of $T_D$ increases, resulting in a decrease
of $\Omega_{\tz_1}h^2$ when re-annihilation dominates the neutralino relic abundance.
%%%%%%%%%%%%%%%%%%%%%%%%%%%%%%%%%%%%%%%%%%%%%%%%%%%%%%%%%%%%%%%%%%%
\FIGURE[t]{
\includegraphics[width=10cm]{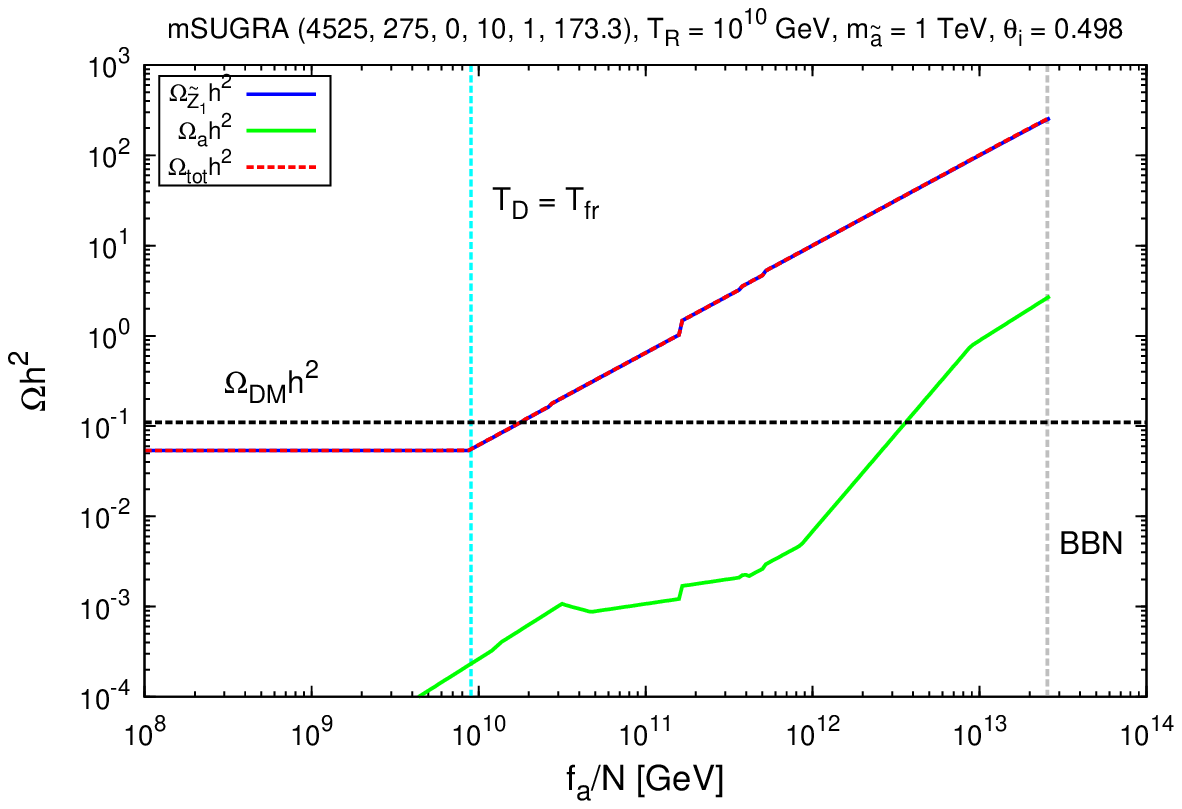}
\includegraphics[width=10cm]{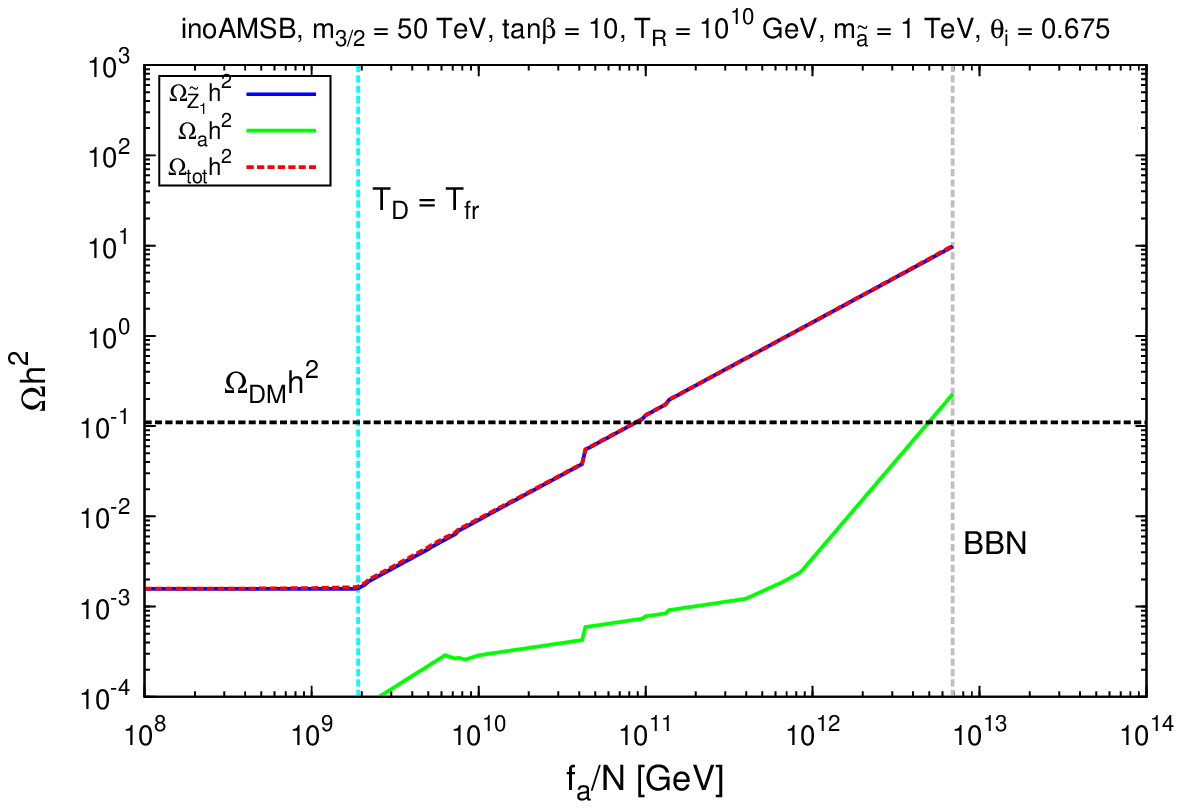}
\caption{Plot of neutralino and axion relic densities  $\Omega h^2$ versus $f_a/N$ for
$m_{\ta}=1$ TeV and $T_R=10^{10}$ GeV
for {\it a}) the HB/FP model and {\it b}) the inoAMSB model.
}\label{fig:Oh2tot_fa}}
%%%%%%%%%%%%%%%%%%%%%%%%%%%%%%%%%%%%%%%%%%%%%%%%%%%%%%%%%%%%%%%%%%%

In Fig. \ref{fig:Oh2tot_tr}, we show the mixed dark matter relic abundance
versus $T_R$ for fixed $f_a/N=10^{12}$ GeV and fixed $m_{\ta}=1$ TeV, for benchmarks
BM1 and BM2. In this case, $T_D$ is fixed throughout the plots, and so 
$\Omega_{\tz_1}h^2$ is nearly constant everywhere except at low $T_R\sim 10^4$ GeV,
where thermal axino production is somewhat suppressed, and fewer neutralinos
are produced at $T_D$ to enter the re-annihilation process. Since $f_a/N$ is
fixed, the axion abundance is also constant throughout much of the plot. 
At $T_R\agt 10^8-10^9$ GeV, we enter the region where axinos can dominate the
universe ($r>1$), and entropy production from axino decay diminishes
the axion abundance. For even higher values of $T_R>T_{dcp}$, the axino
production rate becomes independent of $T_R$, and the entropy injection
ratio $r$ becomes constant with $T_R$. While the neutralino abundance dominates the
axion abundance in these frames, again, this is just a reflection of the
value of $m_{\ta}$ chosen; for higher $m_{\ta}$, $T_D$ will increase, leading
to a diminution of $\Omega_{\tz_1}h^2$. 
%%%%%%%%%%%%%%%%%%%%%%%%%%%%%%%%%%%%%%%%%%%%%%%%%%%%%%%%%%%%%%%%%%%
\FIGURE[t]{
\includegraphics[width=10cm]{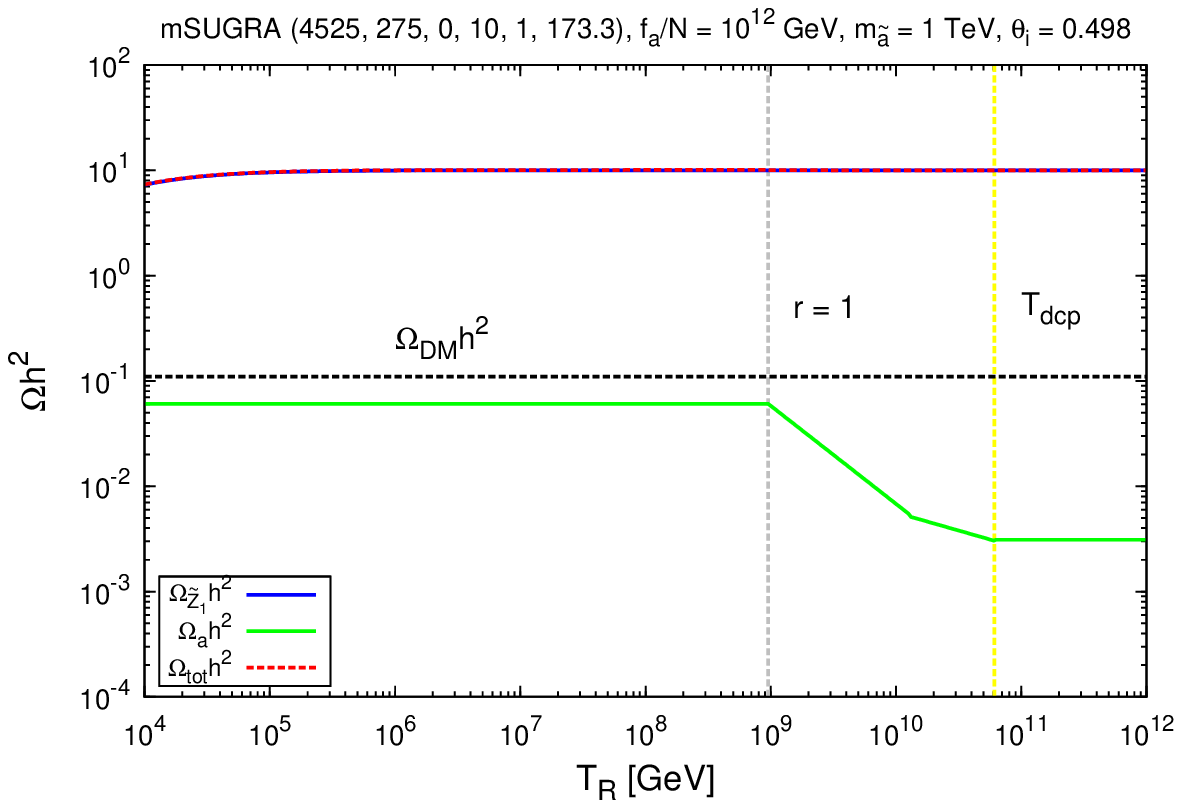}
\includegraphics[width=10cm]{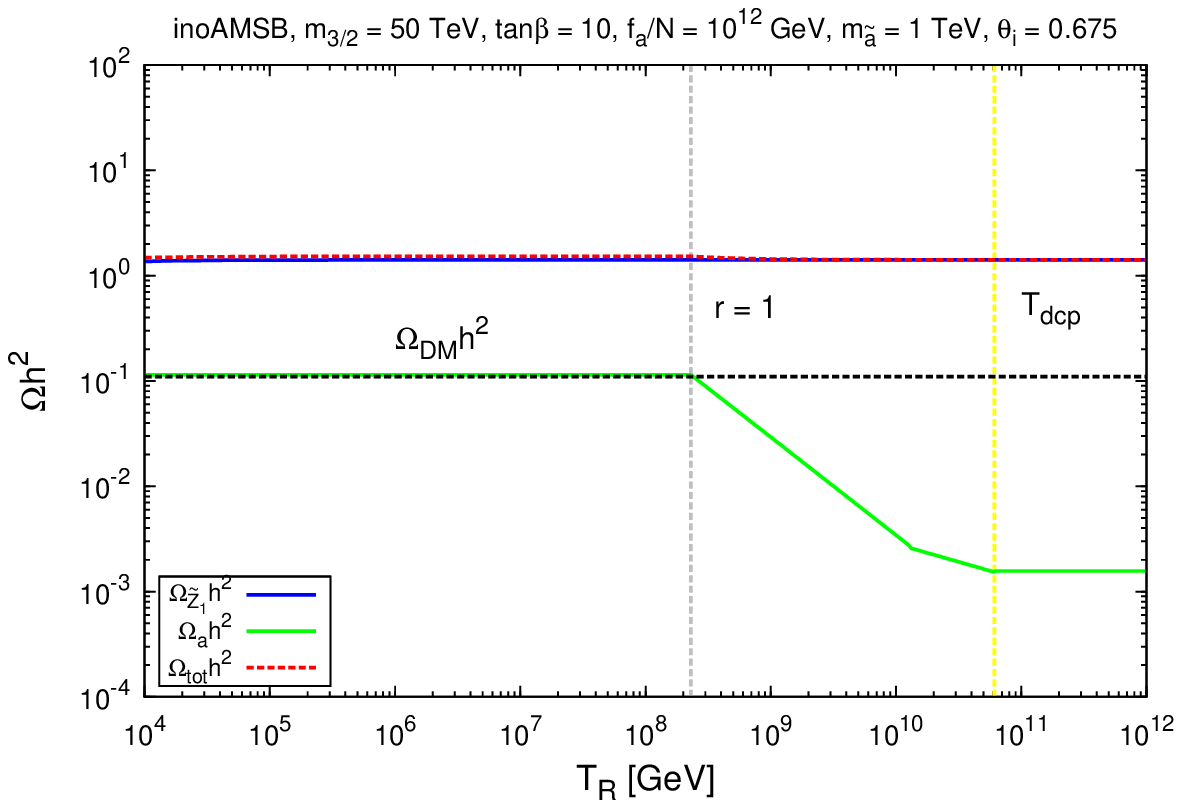}
\caption{Plot of neutralino and axion relic densities  $\Omega h^2$ versus $T_R$ for
$m_{\ta}=1$ TeV and $f_a/N=10^{12}$ GeV
for {\it a}) the HB/FP model and {\it b}) the inoAMSB model.
}\label{fig:Oh2tot_tr}}
%%%%%%%%%%%%%%%%%%%%%%%%%%%%%%%%%%%%%%%%%%%%%%%%%%%%%%%%%%%%%%%%%%%

We have seen that over most of parameter space with $T_D<T_{fr}$, 
$\Omega_{\tz_1}h^2\sim 1/T_D\sim \Gamma_{\ta}^{-1/2}\sim(f_a/N)/m_{\ta}^{3/2}$, with little 
dependence on $T_R$. Hence, a good way to display the relic density of dark 
matter in the mixed $a\tz_1$ CDM scenario is to display it in the $m_{\ta}\ vs.\ (f_a/N)$
plane for benchmarks BM1 and BM2. This plane is shown in Fig. \ref{fig:Oh2tot_fama}{\it a}).
for BM1 and frame {\it b}). for BM2. Here, we take $\theta_i=0.498$ so as to normalize
the relic density $\Omega_{a\tz_1}h^2$ to the measured value $0.1123$ when
$T_D>T_{fr}$ and $f_a/N=10^{12}$ GeV. The black contour denotes the line
where $T_D=T_{fr}$: below and right of this contour, the neutralino 
relic density is given by its usual thermal abundance, which is $\Omega_{\tz_1}h^2=0.05$
for the BM1 case in frame {\it a}). In this region, the axion abundance increases
with increasing $f_a/N$, so that $\Omega_{a\tz_1}h^2=0.1123$ at $f_a/N=10^{12}$ GeV
by design, with a roughly even admixture of mixed higgsino and axion dark matter
in the narrow azure-shaded region. In the region to the left of the $T_D=T_{fr}$
contour, the neutralino abundance rapidly increases, and we have regions of
dominantly WIMP CDM.

In frame {\it b}). for BM2, we again adjust $\theta_i$ so that 
the total relic density equals the measured value for $T_D>T_{fr}$ with 
$f_a/N=10^{12}$ GeV. We see qualitatively similar behavior as in frame {\it a}).
%except now the azure region in accord with WMAP7 measurements is present for both
%$T_D>T_{fr}$ as well as for $T_D<T_{fr}$.  
The azure region for $T_D>T_{fr}$ (right-side of plot)
has $\Omega_{\tz_1}\sim 0.02$, and so is axion-dominated, while the azure region for
$T_D<T_{fr}$ is wino-dominated. In the region with $T_D<T_{fr}$, 
the relic abundance of winos rapidly increases
as we move to smaller $m_{\ta}$ or larger $f_a/N$ values.
%
%%%%%%%%%%%%%%%%%%%%%%%%%%%%%%%%%%%%%%%%%%%%%%%%%%%%%%%%%%%%%%%%%%%
\FIGURE[t]{
\includegraphics[width=10cm]{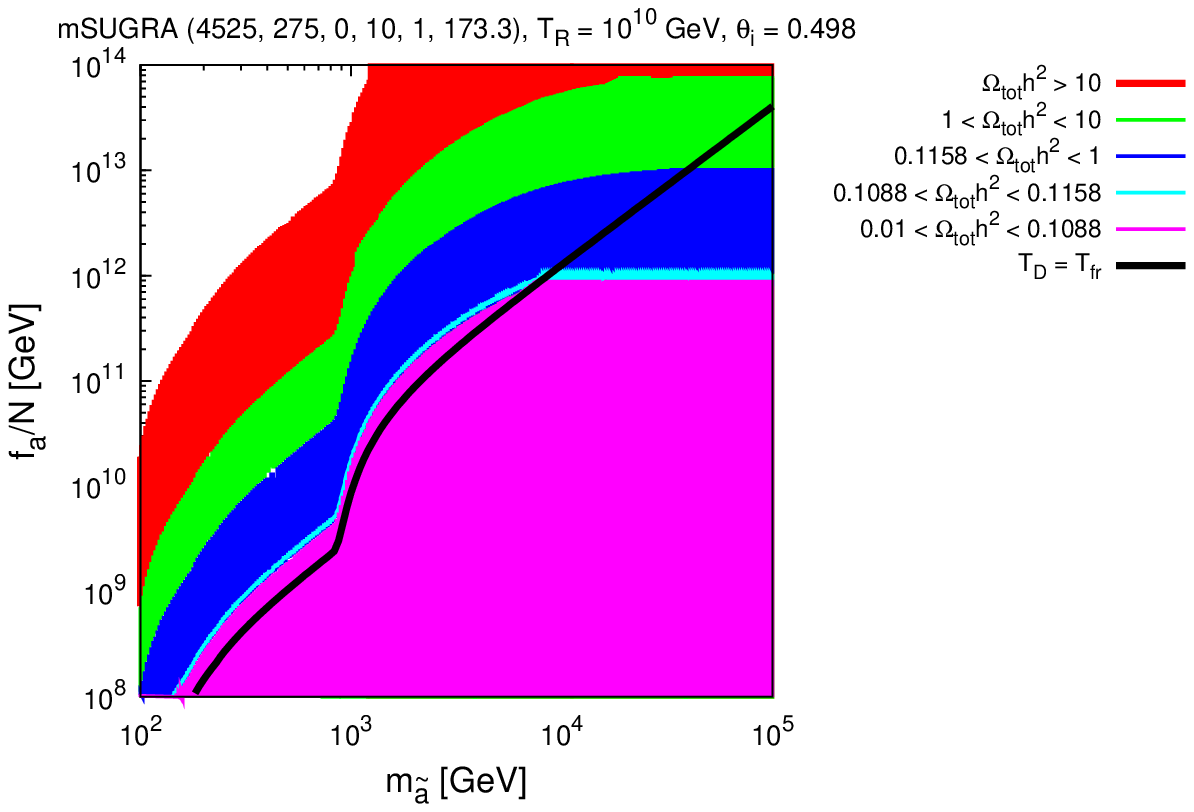}
\includegraphics[width=10cm]{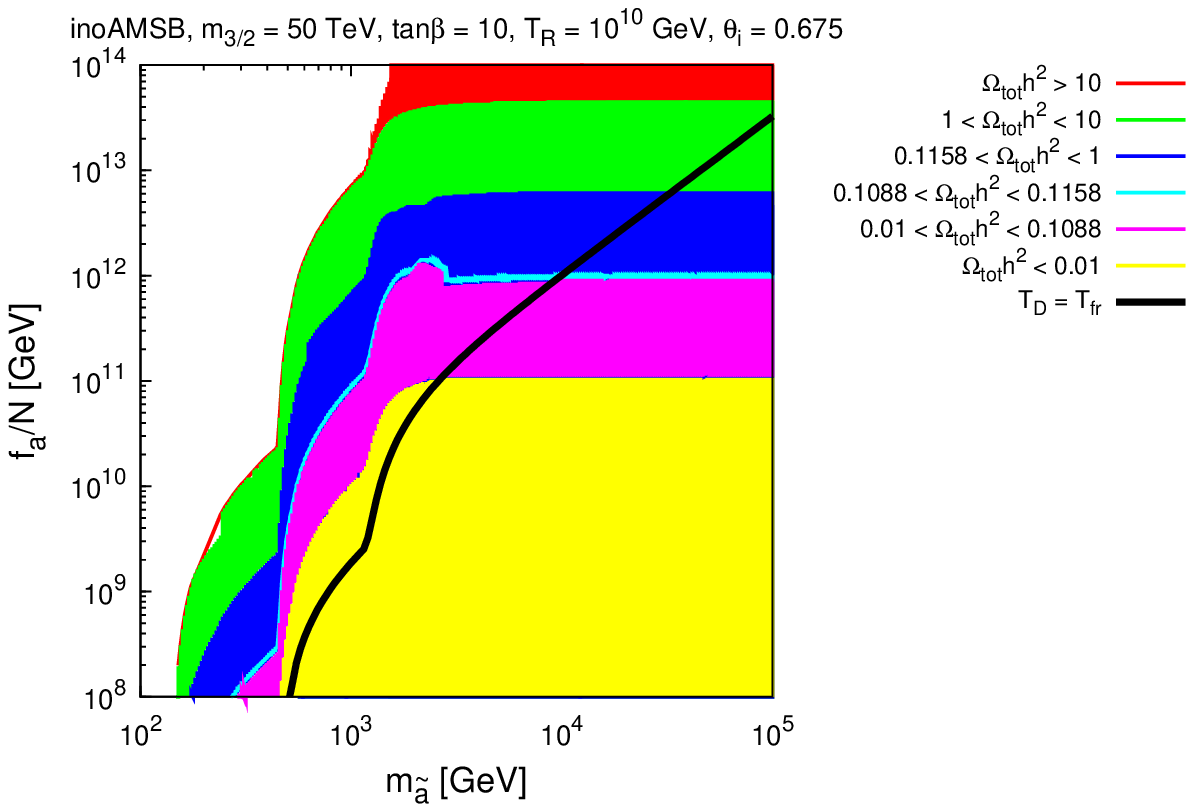}
\caption{Regions of neutralino plus axion relic density  $\Omega_{a\tz_1} h^2$ 
in the $m_{\ta}\ vs.\ f_a/N$ plane for $T_R=10^{10}$ GeV
for {\it a}) the FP model and {\it b}) the inoAMSB model.
The unshaded (white) regions are excluded by BBN bounds since
$T_R<2$ MeV.
}\label{fig:Oh2tot_fama}}
%%%%%%%%%%%%%%%%%%%%%%%%%%%%%%%%%%%%%%%%%%%%%%%%%%%%%%%%%%%%%%%%%%%

From the above results for the benchmarks BM1 and BM2,
we see that the $a\tz_1$ scenario can be classified
into two main cases:
\bi
\item {\it A)}: decoupled axino ($T_D > T_{fr}$)
\item {\it B)}: axino enhanced DM ($T_D < T_{fr}$) .
\ei

Case {\it A)} happens for high $T_D$ values, which are obtained at
low $f_a$ and/or high $m_{\ta}$, as seen in Figs.~\ref{fig:Oh2tot_max}
and \ref{fig:Oh2tot_fa}. In this scenario, the axino has no effect on the
DM relic density, which can be a mixture of axions and neutralinos. Since
the axion mis-alignment angle ($\theta_i$) can always be adjusted so that $\Omega_a h^2 = 0.1123$,
there is no lower bound on $\Omega_{\tz_1} h^2$. Nonetheless the
neutralino relic density must still satisfy:
\be
\Omega_{\tz_1} h^2 = \Omega_{\tz_1}^{std} h^2 \leq 0.1123\;\; (T_D > T_{fr}) \label{eq:DMconst}
\ee
where $\Omega_{\tz_1}^{std} h^2$ is the standard neutralino freeze-out relic density 
in the MSSM, since there is no axino dilution or contribution in this case. Therefore,
in Case A, any MSSM model satisfying Eq. \ref{eq:DMconst} is allowed. For
models where $\Omega_{\tz_1}^{std} h^2 < 0.1123$, the remaining of the DM is composed
of axions.

For Case {\it B)}, $T_D < T_{fr}$, which is obtained at
high $f_a$ and/or low $m_{\ta}$. As shown in Figs. \ref{fig:Y} and \ref{fig:Ym5},
for most of the parameter space considered here, the neutralino relic
density is dominated by the annihilation term in Eq. \ref{eq:reann}. In this case
the relic density can be approximated by:
\be
\Omega_{\tz_1} h^2 \simeq \Omega_{\tz_1}^{std} h^2 \times \frac{T_{fr}}{T_D} .
\ee
Assuming $T_{fr} \sim m_{\tz_1}/20$ and using Eqs. \ref{eq:TD} and \ref{eq:Gammag}, we obtain:
\be
\Omega_{\tz_1} h^2 \simeq 25 \times \Omega_{\tz_1}^{std} h^2 \left(\frac{m_{\tz_1}}{100\ {\rm GeV}}\right) \left(\frac{f_a/N}{10^{12}\ {\rm GeV}}\right) \left(\frac{1\ {\rm TeV}}{m_{\ta}}\right)^{3/2} \left(1-\frac{m_{\tg}^2}{m_{\ta}^2} \right)^{-3/2}
\ee
where we assumed $m_{\ta} \gtrsim m_{\tg}$. Now imposing the
DM relic density constraint (Eq. \ref{eq:DMconst}), we obtain: 
\be
\Omega_{\tz_1}^{std} h^2 \lesssim 4\times10^{-3} \left(\frac{100\ {\rm GeV}}{m_{\tz_1}}\right) \left(\frac{10^{12}\ {\rm GeV}}{f_a/N}\right) \left(\frac{m_{\ta}}{1\ {\rm TeV}}\right)^{3/2} \left(1-\frac{m_{\tg}^2}{m_{\ta}^2} \right)^{3/2}\;\; (T_D < T_{fr}) .
\label{eq:DMconstB}
\ee
Therefore, in this case, the MSSM relic density has to be considerably suppressed
in order to satisfy the above bound. Although the bound decreases with $f_a$ and increases
with $m_{\ta}$, for sufficiently low $f_a$ and/or high $m_{\ta}$, then $T_D > T_{fr}$ and the bound in
Eq. \ref{eq:DMconst} must be used instead. 
In the case where $T_D<T_{fr}$, the DM will likely be composed mainly of
relic neutralinos, unless $\Omega_{\tz_1}^{std} h^2$ is much smaller than Eq. \ref{eq:DMconstB}.
We also point out that the approximate bound in Eq. \ref{eq:DMconstB} is a conservative one,
since, for $m_{\ta} < m_{\tg}$, the bound would be more strict.

%From the above results it is trivial to see that the FP/HB point with $\Omega_{\tz_1}^0 h^2 = 0.05$ 
%can satify the DM constraints if $T_D > T_{fr}$, as shown in Figs. \ref{fig:Oh2tot_max} and \ref{fig:Oh2tot_fa}. 
%On the other hand, the inoAMSB case, which has a smaller $\Omega_{\tz_1}^0 h^2$ value can satisfy the DM 
%constraint even for $T_D < T_{fr}$, as seen in Figs. \ref{fig:Oh2tot_max} and \ref{fig:Oh2tot_fa}.

%From the above results it is trivial to see that even if the FP/HB point with $\Omega_{\tz_1}^0 h^2 = 0.05$
%can satify the DM constraints for both $T_D > T_{fr}$ and $T_D < T_{fr}$, there is not much room for the 
%latter case as shown in Figs. \ref{fig:Oh2tot_max} and \ref{fig:Oh2tot_fa}.
%On the other hand, the inoAMSB case, which has much smaller $\Omega_{\tz_1}^0 h^2$ value can more easily 
%satisfy the DM constraint as seen in Figs. \ref{fig:Oh2tot_max} and \ref{fig:Oh2tot_fa}.

To see how these conclusions depend on the SUSY spectrum, we show in Fig. \ref{fig:inoamsb} the neutralino, axion and
summed relic abundance for the inoAMSB model versus $m_{3/2}$ for
$\tan\beta =10$ and $\mu >0$. 
The results hardly change with varying
$\tan\beta $ or $\mu$. We also take $f_a/N=10^{12}$ GeV, 
$m_{\ta}=2$ TeV, $\theta_i =0.675$ and $T_R=10^{10}$ GeV.
As $m_{3/2}$ increases, all the sparticle masses increase as well and so $\Gamma_{\ta}$ decreases. 
Once $m_{3/2} \simeq 50$~TeV,
$m_{\tg} \simeq m_{\ta}/2$ and the bound in Eq. \ref{eq:DMconstB} can no longer be satisfied.
% and consequently the neutralino abundance, 
%which is proportional to $1/T_D$, increases as well.
The ratio of entropy injection $r$ is shown as the yellow curve, 
against the right-hand $y$-axis. Since $r=4m_{\ta}Y_{\ta}/3 T_D$, 
and $T_D\sim \Gamma_{\ta}^{1/2}$, $T_D$ decreases with increasing $m_{3/2}$,
and the entropy ratio increases. This leads to a dilution of the axion
relic density as shown in the plot.
The jog in the curves around $m_{3/2}\sim 55$ TeV occurs due to a change in
the degrees of freedom $g_*$.
%
%%%%%%%%%%%%%%%%%%%%%%%%%%%%%%%%%%%%%%%%%%%%%%%%%%%%%%%%%%%%%%%%%%%
\FIGURE[t]{
\includegraphics[width=10cm]{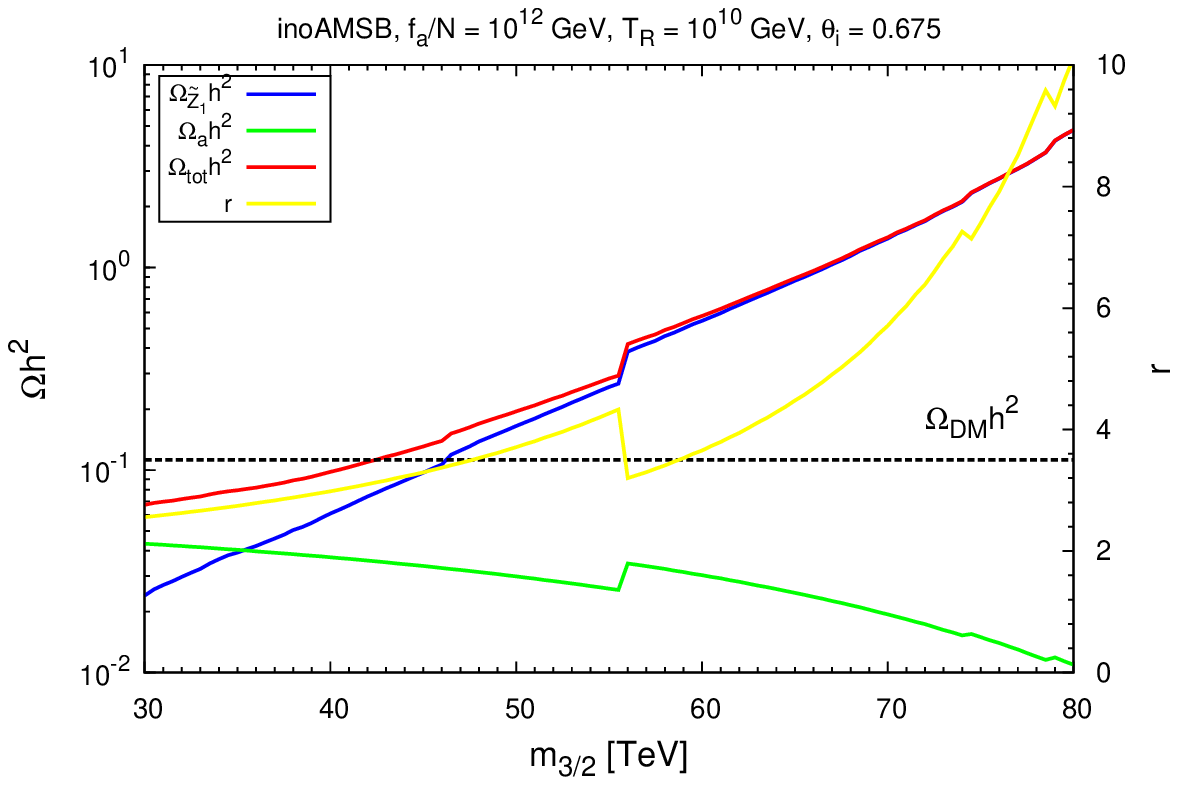}
\caption{Neutralino, axion and total relic density for the inoAMSB
model versus $m_{3/2}$ with $\tan\beta =10$ and $\mu >0$
and for $m_{\ta}=2$ TeV and $\theta_i =0.675$.
We also show the value of $r$ by the yellow curve, and the right-side axis.
}\label{fig:inoamsb}}
%%%%%%%%%%%%%%%%%%%%%%%%%%%%%%%%%%%%%%%%%%%%%%%%%%%%%%%%%%%%%%%%%%%

\section{Effect of saxion production and decay on relic abundance}
\label{sec:saxion}

We have seen so far that the relic neutralino abundance may be 
enhanced beyond usual expectations in the mixed $a\tz_1$ scenario if
$T_D<T_{fr}$. However, we have so far neglected a mandatory element
of the axion supermultiplet: the spin-0, $R$-parity even saxion field $s$\cite{saxions}.
Saxions may be produced thermally in the early universe, either in
thermal equilibrium for $T_R>T_{dcp}$ or via radiation and decay for
$T_R<T_{dcp}$. Saxions can also be produced via coherent oscillations.

In an analagous manner to the axino case, the saxion field can dominate the energy density of the universe
at early times, if its decay temperature ($T_{D_s}$) is smaller than the temperature ($T_{e_s}$)
at which its energy density overcomes the radiation and axino energy densities.
The saxion can decay into gluons and gluinos (and perhaps axions), depending on its mass and its (model dependent)
couplings. As in the axino case, the saxion decay will inject entropy at $T=T_{D_s}$, 
diluting the axino (if $T_D < T_{D_s}$),
neutralino (if $T_{fr} > T_{D_s}$) and axion (if $T_a > T_{D_s}$) relic densities.
Furthermore, if the $s\to \tg \tg$ branching ratio is considerable, saxion decays will also inject neutralinos
through gluino cascade decays. In the case where a high rate of neutralino injection occurs after freeze-out,
one must again consider a second possibility of neutralino 
re-annihilation which may enhance the neutralino abundance.

%The saxion decay width $\Gamma_s$, 
%assuming dominant decays $s\to gg$ and $s\to\tg\tg$,
%has been computed in Ref. \cite{ay}. In an analagous manner to the axino
%case, the re-heat temperature associated with saxion decay can be computed
%as $T_{D_s}$. 
%If $T_{D_s}\agt T_{fr}$, then saxion decay products will thermalize before
%neutralino freeze-out, and all the previous results will be applicable.
%For $f_a/N\sim 10^{12}$ GeV, this occurs for $m_s\agt 2.5$ TeV.
%If $T_{D_s}<T_{fr}$, then the temperature $T_{s=rad}$ may also be 
%computed to see if there
%is a temperature at which saxions dominate the universe. If saxions dominate 
%and if $T_{D_s}<T_{s=rad}$, then saxions can inject considerable entropy
%$r_s$ and dilute whatever frozen-out relics exist at $T=T_{D_s}$. 
%Contours of $r_s$ are shown in the $f_a/N\ vs.\ T_R$ plane for several
%values of $m_s$ in Ref. \cite{ay}.
%Entropy injection from saxion production and decay thus has the chance to 
%diminish the neutralino relic abundance in models where the 
%neutralino is the LSP. In addition, saxion decay to final states such as
%$\tg\tg$ have the potential to inject additonal neutralinos into the
%cosmic soup, either before or after the usual neutralino freeze-out. 
%In the case where a high rate of neutralino injection occurs after freeze-out,
%$Y_{\tz_1}^s>Y_{\tz_1}^{th}(T=T_{D_s})$, 
%then again one must consider a second possibility of neutralino 
%re-annihilation which may enhance the neutralino abundance. 

Since the saxion lifetime is comparable to the axino lifetime, it is possible 
that both saxions and axinos may co-dominate the universe. This makes a simplistic
analysis of neutralino abundance difficult using the approach of this paper.
To make matters worse, gravitinos may be produced thermally at significant 
rates for high enough $T_R$, and enjoy decays at time scales comparable
to heavy axinos and saxions.
Thus, gravitino production and decay may also enhance or diminish the neutralino
abundance.

The proper treatment of such intricately coupled effects is best made by 
numerical solution of the coupled Boltzmann equations. This type of treatment has
been initiated in Ref. \cite{blrs2}, and will be reported at a future date.

%==============================================================================
\section{Conclusions}
\label{sec:conclude}
%==============================================================================

In this paper, we have presented results of calculations of the dark matter
abundance in supersymmetric models wherein the strong $CP$ problem is solved by
the Peccei-Quinn mechanism, and in which the neutralino is the LSP, so that 
dark matter consists of an axion/neutralino admixture. Since the $a\tz_1$ CDM
scenario boosts the dark matter abundance beyond the usual thermal neutralino production rates, 
we have presented results for two models that typically yield an underabundance
of thermal neutralino dark matter: the HB/FP region of mSUGRA with a mixed
higgsino-like neutralino and AMSB-type models with a wino-like neutralino. 
Our final results depend mainly on the temperature $T_D$ at which heavy axinos
finish their cascade decays to neutralinos. 

In the case where $T_D>T_{fr}$, the neutralino abundance is given by its usual
thermal abundance. 
In the case where there is an underabundance of neutralinos compared to the measured
dark matter abundance, then the remainder can be comprised of axions. In the case of our two 
benchmark models, in the HB/FP region, we would obtain a nearly equal mixture 
of axions and mixed higgsino-like neutralinos, while in the BM2 case of the 
inoAMSB model, we would obtain a case with mainly axion CDM, along with a small
admixture of wino-like neutralinos. The case of BM1 provides an instance where
both WIMP and axion signals\cite{admx} could ultimately be found at dark matter detectors.
In the case of BM2, only an axion signal might be discovered. The direct detection rates for
wino-like WIMPs has been presented in Ref. \cite{shibi}. These projected rates would have to
be scaled down by a factor of $\sim 70$ since in this case wino-like WIMPs would
only comprise $\sim 1/70$ of the total dark matter abundance.

In the case where $T_D<T_{fr}$, heavy axino decays in the early universe will
inject additional neutralinos and entropy into the cosmic soup after neutralino
freeze-out. The main effect here is that neutralino re-annihilation takes place at
$T=T_D$, and one obtains a neutralino abundance as if the freeze-out temperature
were replaced by $T_D$. In this case, $\Omega_{\tz_1}h^2\sim 1/T_D$ instead of
$1/T_{fr}$, and since $T_D<T_{fr}$, one obtains a greatly enhanced abundance of
neutralinos beyond the usual expectation. The neutralino ``production by decay and 
re-annihilation'' mechanism can thus lead to enhanced production of
wino-like or higggsino-like neutralinos-- which naively give rise to an underabundance of
dark matter-- so that $\tz_1$s might comprise nearly all the abundance of dark matter.
Also in this case, if $T_e>T_D$, then significant entropy production by
decaying axinos can diminish the axion abundance. In this case, one might
expect mainly wino-like or higgsino-like dark matter, with a small
admixture of axions. Thus, the enhanced neutralino production via axino decay mechanism
offers an alternative means to allow wino- or higgsino-like neutralinos to comprise
the bulk of dark matter. In many respects, this mechanism may be preferred over
the possibility of multi-TeV scale moduli field decay\cite{moroi,kane}, since it also
allows for a solution to the strong $CP$ problem.

%==============================================================================
\acknowledgments
%==============================================================================

This research was supported in part by the U.S. Department of Energy,
by the Fulbright Program and CAPES (Brazilian Federal Agency for
Post-Graduate Education).

\appendix

\section{Expansion rate of early universe}
\label{app:exrate}

Here we briefly review the cosmology of an early axino dominated universe and in the
next Section present
the expressions for the axion relic density and neutralino yield used in Secs.~\ref{sec:z1}-\ref{sec:results}.

First, we define several temperatures:
\bi 
\item $T_e$: temperature when the universe becomes axino (matter) dominated,
\item $T_S$: temperature at which entropy injection due to axino decay starts,
\item $T_D$: temperature at which entropy injection due to axino decay ends.
\ei

\subsection{Matter dominated phase: $T_S < T < T_e$}

In this phase, the universe is axino dominated, for which:
\begin{eqnarray}
\rho_{\ta} & = & \rho_0 \frac{R_e^3}{R^3} \exp(-(t-t_e)\Gamma_{\ta}) \approx \rho_0 \frac{R_e^3}{R^3} \\
%\rho_{\gamma} & = & \frac{\pi^2}{30} g_*(T) T^4\\
H(T) & = & \sqrt{\frac{\rho_{\ta}}{3 M_P^2}} \simeq \sqrt{\frac{\rho_0}{3 M_P^2}}\left(\frac{R_e}{R}\right)^{3/2}
\end{eqnarray}
where $R_e = R(T_e)$ and $\Gamma_{\ta}$, $\rho_{\ta}$, $Y_{\ta}$ and $m_{\ta}$ are the axino width, energy density,
yield and mass, respectively. 
Here, $M_{P}=M_{Pl}/\sqrt{8\pi}$ ({\it i.e.} $M_P$ is the reduced Planck mass), and we 
will use $g_{*S}=g_*$ since the temperatures we consider are all above $T_{BBN}\simeq 2$ MeV.
By definition we have:
\be
\rho_0 = \frac{\pi^2}{30} g_*(T_e) T_e^4\; \mbox{ and } T_e = \frac{4}{3} Y_{\ta} m_{\ta} .
\ee
 Since entropy is still conserved in this phase,
\be
\frac{g_*(T) T^3}{g_*(T_e) T_e^3} = \left(\frac{R_e}{R}\right)^3 ,
\ee
and hence
\be
H(T) = \sqrt{\frac{\pi^2}{90} g_*(T) T_e} \frac{T^{3/2}}{M_P} .
\ee

\subsection{Decaying particle dominated phase: $T_D < T < T_S$}

In this phase, the universe is dominated by a decaying particle\cite{mcdonald}, which gives:
\begin{eqnarray}
\rho_{\ta} & = & \rho_0 \frac{R_e^3}{R^3} \exp(-(t-t_e)\Gamma_{\ta}) \approx \rho_0 \frac{R_e^3}{R^3}\\
%\rho_{\gamma} & = & \frac{\pi^2}{30} g_*(T) T^4\\
H(T) & = & \sqrt{\frac{\rho_{\ta}}{3 M_P^2}} \approx \sqrt{\frac{\rho_0}{3 M_P^2}}\left(\frac{R_e}{R}\right)^{3/2} .
\end{eqnarray}
Entropy is no longer conserved, so that\cite{scherrer,lps}:
\be
\frac{g_*(T)^2 T^8}{g_*(T_0)^2 T_0^8} = \left(\frac{R_0}{R}\right)^3 .
\ee
Using this relation, we obtain:
\be
H(T) = H(T_D) \left(\frac{R(T_D)}{R(T)}\right)^{3/2} = H(T_D) \frac{g_*(T) T^4}{g_*(T_D) T_D^4} .
\ee
But, at $T=T_D$ all the matter energy has been converted to radiation, hence
\begin{eqnarray}
H(T_D) & = & \sqrt{\frac{\pi^2}{90} g_*(T_D)}\frac{T_D^2}{M_P} \ \ \ {\rm or}\\
H(T) & = & \sqrt{\frac{\pi^2}{90}}\frac{g_*(T)}{\sqrt{g_*(T_D)}}\frac{T^4}{T_D^2 M_P}
\end{eqnarray}
The decay temperature ($T_D$) and $\Gamma_{\ta}$ are related by $H(T_D) = \Gamma_{\ta}$, so that 
\be
\Gamma_{\ta}^2  =  \frac{\pi^2}{90} g_*(T_D) \frac{T_D^4}{M_P^2} .
\ee

\subsection{Radiation dominated phase: $T < T_D$}

In this phase the universe is radiation dominated, giving the standard expressions:
\be
H(T) =  \sqrt{\frac{\rho_{\gamma}}{3 M_P^2}} = \sqrt{\frac{\pi^2}{90} g_*(T)} \frac{T^2}{M_P}
\ee
and entropy is once again conserved:
\be
\frac{g_*(T) T^3}{g_*(T_0) T_0^3} = \left(\frac{R_0}{R}\right)^3 .
\ee

\section{Axion oscillation}
\label{app:AO}

The axion field starts to oscillate when\footnote{Here, we follow much of the notation given by Visinelli and Gondolo\cite{vg2}.}
\be
3H(T_a) = m_a(T_a)
\ee
where the temperature-dependent axion mass is given by
\be
m_a(T) = \left\{ \begin{array}{ll}
 m_a \mbox{, if $T < \Lambda$} \\
m_a b \left(\frac{\Lambda}{T}\right)^4 \mbox{, if $T > \Lambda$} \end{array} \right.
\ee
with $b = 0.018$, $\Lambda = 0.2$ GeV and $m_a = 6.2\times10^{-3}/f_a$. 
Due to its temperature dependent mass, after oscillation begins, the axion energy density obeys
\be
\frac{\rho_a(T) R(T)^3}{m_a(T)} = constant .
\ee

If $T_a < T_D$, then the axion field starts to oscillate after the axino has decayed.
In this case, the axion relic density is given by the standard expression\cite{vg1,vg2}:
\be
\Omega_a^{std} h^2 = \left\{ \begin{array}{ll}
 9.23\times10^{-3} \theta_i^2 f(\theta_i)  \frac{1}{g_*(T_a)^{1/4}} \left(\frac{f_a}{10^{12}}\right)^{3/2} \mbox{, if $f_a > \hat{f}_a$} \\
 1.32\ \theta_i^2 f(\theta_i) \frac{1}{g_*(T_a)^{5/12}}\left(\frac{f_a}{10^{12}}\right)^{7/6} \mbox{, if $f_a < \hat{f}_a$} \label{case0}
\end{array} \right.
\ee
where $\hat{f}_a = 9.9\times10^{16}$ GeV, $f(\theta_i) = \ln(\frac{e}{1-\theta_i^2/\pi^2})^{7/6}$ 
and the standard oscillation  temperature is given by
\be
T_a^{std} = \left\{ \begin{array}{ll}
1.23\times 10^2  \frac{1}{g_*(T_a)^{1/4}} \left(\frac{10^{12}}{f_a}\right)^{1/2} \mbox{, if $f_a > \hat{f}_a$} \\
8.71\times10^{-1} \frac{1}{g_*(T_a)^{1/12}} \left(\frac{10^{12}}{f_a}\right)^{1/6} \mbox{, if $f_a < \hat{f}_a$}
\end{array} \right. 
\ee

If instead $T_e < T_a$, then the axion density is diluted by the entropy ratio $r$ so that
\be
\Omega_a h^2=\frac{1}{r}\times\Omega_a^{std}h^2
\ee

If $T_D < T_a < T_e$, the axion can start to oscillate in the matter dominated phase (MD) or the decaying dominate phase (DD). 
The relic densities for each case are:
\bi
\item Matter dominated ($T_S < T_a < T_e$):
\ei
\be
\Omega_a^{MD} h^2 = \left\{ \begin{array}{ll}
 7.5\times10^{-5}\ \theta_i^2 f(\theta_i) T_D \left(\frac{f_a}{10^{12}}\right)^2 \mbox{, if $f_a > \hat{f}_a^{MD}$} \\
 1.4\ \theta_i^2 f(\theta_i) \frac{1}{g_*(T_a)^{4/11}}\left(\frac{f_a}{10^{12}}\right)^{14/11} \frac{T_D}{T_e^{4/11}} \mbox{, if $f_a < \hat{f}_a^{MD}$}
\end{array} \right.
\ee
with
\be
\hat{f}_a^{MD} =  7.6\times10^{17} \frac{1}{\sqrt{g_*(T_a) T_e}}
\ee
and $T_a$ given by

\be
T_a^{MD} = \left\{ \begin{array}{ll}
%\left( \sqrt{\frac{5}{4\pi^3}} \frac{1}{\sqrt{g_*(T_a) T_e}} m_a M_{P} \right)^{2/3}  \mbox{, if $f_a > \hat{f}_a^{(A)}$} \\
%\left( b \Lambda^4 \sqrt{\frac{5}{4\pi^3}} \frac{1}{\sqrt{g_*(T_a) T_e}} m_a M_{P} \right)^{2/11}  \mbox{, if $f_a < \hat{f}_a^{(A)}$}
6.1\times10^2 \left(\frac{1}{\sqrt{g_*(T_a) T_e}} \frac{10^{12}}{f_a} \right)^{2/3}  \mbox{, if $f_a > \hat{f}_a^{MD}$} \\
8.6\times10^{-1} \left(\frac{1}{\sqrt{g_*(T_a) T_e}} \frac{10^{12}}{f_a} \right)^{2/11}  \mbox{, if $f_a < \hat{f}_a^{MD}$}

\end{array} \right. .
\ee

\bi
\item Decaying particle dominated ($T_D < T_a < T_S$)\cite{vg2}:
\ei
\be
\Omega_a^{DD} h^2 = \left\{ \begin{array}{ll}
 7.5\times10^{-5}\ \theta_i^2 f(\theta_i) T_D \left(\frac{f_a}{10^{12}}\right)^2 \mbox{, if $f_a > \hat{f}_a^{DD}$} \\
 1.72\ \theta_i^2 f(\theta_i) \frac{g_*(T_D)^{1/4}}{\sqrt{g_*(T_a)}} T_D^2 \left(\frac{f_a}{10^{12}}\right)^{3/2} \mbox{, if $f_a < \hat{f}_a^{DD}$}
\end{array} \right.
\ee
with
\be
\hat{f}_a^{DD} =  5.69\times10^{20} \frac{\sqrt{g_*(T_D)}}{g_*(T_a)} T_{D}^2
\ee
and $T_a$ given by:
\be
T_a^{DD} = \left\{ \begin{array}{ll}
%\left( \sqrt{\frac{5}{4\pi^3}} \frac{\sqrt{g_*(T_D)}}{g_*(T_a)} m_a M_{P} T_D^2 \right)^{1/4}  \mbox{, if $f_a > \hat{f}_a^{(B)}$} \\
%\left( b\Lambda^4 \sqrt{\frac{5}{4\pi^3}} \frac{\sqrt{g_*(T_D)}}{g_*(T_a)} m_a M_{P} T_D^2 \right)^{1/8} \mbox{, if $f_a < \hat{f}_a^{(B)}$}
0.11\times10^2 \left(\frac{\sqrt{g_*(T_D)}}{g_*(T_a)} \frac{10^{12}}{f_a} T_D^2 \right)^{1/4}  \mbox{, if $f_a > \hat{f}_a^{DD}$} \\
9.0\times10^{-1} \left(\frac{\sqrt{g_*(T_D)}}{g_*(T_a)} \frac{10^{12}}{f_a} T_D^2 \right)^{1/8} \mbox{, if $f_a < \hat{f}_a^{DD}$}
\end{array} \right. .
\ee

Matching both solutions we have:
\begin{eqnarray*}
\left(\frac{f_a}{10^{12}}\right)^{5/22} & < & 0.8 \frac{g_*(\bar{T}_a)^{3/22}}{g_*(T_D)^{1/4}} \frac{1}{T_D T_e^{4/11}}  \;\rightarrow \mbox{ MD case} \\
\left(\frac{f_a}{10^{12}}\right)^{5/22} & > & 0.8 \frac{g_*(\bar{T}_a)^{3/22}}{g_*(T_D)^{1/4}} \frac{1}{T_D T_e^{4/11}} \;\rightarrow \mbox{ DD case}
\end{eqnarray*}
or in terms of $T_a$:
\begin{eqnarray*}
T_a & > & T_{S}\;\rightarrow \mbox{ MD case} \\
T_a & < & T_{S}\;\rightarrow \mbox{ DD case} \\
\end{eqnarray*}
where
\be
T_{S} = \left(\frac{g_*(T_D)}{g_*(T_a^{(A)})} T_e T_D^4\right)^{1/5}
\ee

To summarize:
\be
\Omega_ah^2 = \left\{ \begin{array}{ll}
\Omega_a^{std}h^2/r \;  \mbox{, if $T_e < T_a$} \\
\Omega_a^{MD}h^2\;  \mbox{, if $T_{S} < T_a < T_e$} \\
\Omega_a^{DD}h^2\;  \mbox{, if $T_D < T_a < T_{S}$} \\
\Omega_a^{std}h^2 \;  \mbox{, if $T_a < T_D$}
\end{array} \right. ,
\label{eq:Oh2axion}
\ee
where $r$ is the entropy injection ratio as usual.

\section{Neutralino yield}
\label{app:z1}

The neutralino will decouple from the thermal bath when
\be
\langle \sigma v\rangle n_{\tz_1}(T_{fr}) = H(T_{fr})
\ee
where
\be
n_{\tz_1}(T) = 2 \left(\frac{m T}{2 \pi}\right)^{3/2} e^{-m/T} .
\ee
In a radiation dominated universe, the neutralino yield is given by:
\be
Y_{\tz_1}(T_{fr}) = \frac{H(T_{fr})}{\langle \sigma v\rangle s(T_{fr})} ,
\ee
while in a matter dominated universe:
\be
Y_{\tz_1}(T_{fr}) = \frac{3}{2} \frac{H(T_{fr})}{\langle \sigma v\rangle s(T_{fr})} .
\ee

As in the axion case, the neutralino can freeze-out before the universe 
becomes matter dominated ($T_{fr} > T_e$), during the MD phase 
($T_S < T_{fr} < T_e$), during the DD phase ($T_D < T_{fr} < T_S$) 
or during the radiation dominated phase ($T_{fr} < T_D$). The neutralino yields
for each of these scenarios are listed below.
\bi
\item Standard case ($T_{fr} < T_D$):
\ei
\be
Y_{\tz_1}^{std}(T_{fr}) = \frac{\left( 90/\pi^2 g_*(T_{fr})\right)^{1/2}}{4\langle\sigma 
v\rangle M_PT_{fr}}
\label{eq:Yz1std}
\ee
where the freeze-out temperature is given by
\be
T_{fr}^{std} = m_{\tz_1}/\ln[\frac{3\sqrt{5}\langle \sigma v\rangle M_P m_{\tz_1}^{3/2}}
{\pi^{5/2}T_{fr}^{1/2}g_*^{1/2}(T_{fr})}] .
\label{tfrz1}
\ee

\bi
\item MD case ($T_S < T_{fr} < T_e$):
\ei
\be
Y_{\tz_1}^{MD} (T_D)=\frac{3}{2} Y_{\tz_1}^{std}(T_{fr}^{MD})\frac{T_D}{\sqrt{T_e T_{fr}^{MD}}}
\ee
where the freeze-out temperature $T_{fr}^{MD}$ is given by:
\be
T_{fr}^{MD} = m_{\tz_1}/\ln[\frac{3\sqrt{5}\langle \sigma v\rangle M_P m_{\tz_1}^{3/2}}
{\pi^{5/2} g_*^{1/2}(T_{fr}^{MD}) T_e^{1/2}}] .
\ee

\bi
\item DD case ($T_D < T_{fr} < T_S$):
\ei
\be
Y_{\tz_1}^{DD}(T_D) = \frac{3}{2}Y_{\tz_1}^{std}(T_{fr}^{DD})
\frac{g_*^{1/2}(T_D)}{g_*^{1/2}(T_{fr}^{DD})} \left(\frac{T_D}{T_{fr}^{DD}}\right)^3
\ee
where the freeze-out temperature is given by
\be
T_{fr}^{DD} = m_{\tz_1}/\ln [\frac{3\sqrt{5}\langle \sigma v\rangle M_P m_{\tz_1}^{3/2}
g_*^{1/2}(T_D)T_D^2}{\pi^{5/2}g_*(T_{fr}^{DD})(T_{fr}^{DD})^{5/2}} ] .
\ee

\bi
\item Case when $T_{fr}>T_e$:
\ei
\be
Y_{\tz_1}(T_D) = Y_{\tz_1}^{std}(T_{fr}^{std})/r= Y_{\tz_1}^{std}(T_{fr}^{std})
\times\frac{T_e}{T_D} .
\ee

To summarize:
\be
Y_{\tz_1}(T_D) = \left\{ \begin{array}{ll}
Y_{\tz_1}^{std}(T_{fr}^{std})/r \;  \mbox{, if $T_e < T_{fr}$} \\
Y_{\tz_1}^{MD}\;  \mbox{, if $T_{S} < T_{fr} < T_e$} \\
Y_{\tz_1}^{DD}\;  \mbox{, if $T_D < T_{fr} < T_{S}$} \\
Y_{\tz_1}^{std}(T_{fr}^{std})\;  \mbox{, if $T_{fr} < T_D$}
\end{array} \right.
\label{eq:Yrgt1}
\ee
where
\be
T_{S} = \left(\frac{g_*(T_D)}{g_*(T_{fr}^{(A)})} T_D^4 T_e\right)^{1/5} .
\ee

% ---- Bibliography ----
%


\begin{thebibliography}{99}
%
\bibitem{wss} H.~Baer and X.~Tata, {\it Weak Scale Supersymmetry: From 
Superfields to Scattering Events}, 
(Cambridge University Press, 2006).
%
\bibitem{bblt} H. Baer, V. Barger, A. Lessa and X. Tata, 
\jhep{1006}{2010}{102}.
%
\bibitem{thooft} G. 't Hooft, \prl{37}{1976}{8}.
%
\bibitem{axreviews} For a recent review, 
see P. Sikivie, \hepph{0509198}; M. Turner, \prep{197}{1990}{67}; 
J. E. Kim and G. Carosi, \rmp{82}{2010}{557}.
%
\bibitem{arg} For a review, see J. E. Kim, \prep{150}{1987}{1}.
%
\bibitem{nedm} For a comprehensive discussion, see 
  M.~Pospelov and A.~Ritz, Annals Phys.\  {\bf 318} (2005) 119.
%
\bibitem{pq} R. Peccei and H. Quinn, \prl{38}{1977}{1440} and \prd{16}{1977}{1791}.
%
\bibitem{ww} S. Weinberg, \prl{40}{1978}{223};
F. Wilczek, \prl{40}{1978}{279}.
%
\bibitem{ksvz} J. E. Kim, \prl{43}{1979}{103};
M. A. Shifman, A. Vainstein and V. I. Zakharov, \npb{166}{1980}{493}.
%
\bibitem{dfsz} M. Dine, W. Fischler and M. Srednicki, \plb{104}{1981}{199};
A. P. Zhitnitskii, \sjp{31}{1980}{260}.
%
\bibitem{astro} D. Dicus, E. Kolb, V. Teplitz and R. Wagoner, \prd{18}{1978}{1829}
and \prd{22}{1980}{839}; for a review, see G. Raffeldt, \hepph{0611350}.
%
\bibitem{pqsusy} H. P. Nilles and S. Raby, \npb{198}{1982}{102};
J. E. Kim and H. P. Nilles, \plb{138}{1984}{150}.
J. E. Kim, \plb{136}{1984}{378}.
%
\bibitem{witten} P. Svrcek and E. Witten, \jhep{0606}{2006}{051}.
%
\bibitem{rtw} K. Rajagopal, M. Turner and F. Wilczek, 
\npb{358}{1991}{447}.
%
\bibitem{ckkr} L. Covi, J. E. Kim and L. Roszkowski, \prl{82}{1999}{4180}; 
L. Covi, H. B. Kim, J. E. Kim and L. Roszkowski, \jhep{0105}{2001}{033}.
%
\bibitem{axinorev} For recent reviews of axino dark matter, see
F. Steffen, \epjc{59}{2009}{557}; L. Covi and J. E. Kim, 
\njp{11}{2009}{105003}.
%
\bibitem{axinoDM} H. Baer, A. Box and H. Summy, 
\jhep{0908}{2009}{080}.
%
\bibitem{ckls} K-Y. Choi, J. E. Kim, H. M. Lee and O. Seto, \prd{77}{2008}{123501}.
%
\bibitem{wmap7} E. Komatsu {\it et al.} (WMAP collaboration), 
arXiv:1001.4538 (2010).
%
\bibitem{msugra} For a recent review, see
R. Arnowitt and P. Nath, arXiv:0912.2273 (2009)
%
\bibitem{hb_fp} K.~L.~Chan, U.~Chattopadhyay and P.~Nath, \prd{58}{1998}{096004};
J.~Feng, K.~Matchev and T.~Moroi, \prl{84}{2000}{2322} and 
\prd{61}{2000}{075005}; see also 
H.~Baer, C.~H.~Chen, F.~Paige and X.~Tata, \prd{52}{1995}{2746} and 
\prd{53}{1996}{6241}; 
H.~Baer, C.~H.~Chen, M.~Drees, F.~Paige and X.~Tata, \prd{59}{1999}{055014}; 
for a model-independent approach, see
H.~Baer, T.~Krupovnickas, S.~Profumo and P.~Ullio, \jhep{0510}{2005}{020}.
%
\bibitem{amsb} L. Randall and R. Sundrum, \npb{557}{1999}{79};
G. Giudice, M. Luty, H. Murayama and R. Rattazzi, \jhep{9812}{1998}{027}.
%
\bibitem{shanta} S. P. de Alwis, \jhep{1003}{2010}{078}.
%
\bibitem{inoamsb} H. Baer, S. P. de Alwis, K. Givens, S. Rajagopalan and H. Summy,
\jhep{1005}{2010}{069};
H. Baer, S. P. de Alwis, K. Givens, S. Rajagopalan and W. Sreethawong,
\jhep{1101}{2011}{005}.
%
\bibitem{axdm} H. Baer and A. Box, \epjc{68}{2010}{523};
H. Baer, A. Box and H. Summy, \jhep{1010}{2010}{023}.
%
\bibitem{shibi} H. Baer, R. Dermisek, S. Rajagopalan and H. Summy, 
JCAP{\bf 1007} (2010) 014.
%
\bibitem{isasug} ISAJET, by H.~Baer, F.~Paige, S.~Protopopescu and
X.~Tata, \hepph{0312045}; see also
H.~Baer, J.~Ferrandis, S.~Kraml and W.~Porod, \prd{73}{2006}{015010}.
%
\bibitem{isajet} F. Paige, S. Protopopescu, H. Baer and X. Tata, \hepph{0312045}; 
http://www.nhn.ou.edu/$\sim$isajet/
%
\bibitem{isared} H. Baer, C. Balazs and A.Belyaev, \jhep{0203}{2002}{042}.
%
\bibitem{steffen} A. Brandenburg and F. Steffen, JCAP{\bf 0408} (2004) 008.
%
\bibitem{strumia} A. Strumia, \jhep{1006}{2010}{036}.
%
\bibitem{scherrer} R. J. Scherrer and M. S. Turner, \prd{31}{1985}{681}.
%
\bibitem{vacmis} L. F. Abbott and P. Sikivie, \plb{120}{1983}{133};
J. Preskill, M. Wise and F. Wilczek, \plb{120}{1983}{127};
M. Dine and W. Fischler, \plb{120}{1983}{137};
M. Turner, \prd{33}{1986}{889}.
%
\bibitem{vg1} L. Visinelli and P. Gondolo, \prd{80}{2009}{035024}.
%
\bibitem{jedamzik} K. Jedamzik, \prd{70}{2004}{063524}
and \prd{74}{2006}{103509}.
%
\bibitem{moroi} T. Moroi and L. Randall, \npb{570}{2000}{455};
B. Acharya, G. Kane, S. Watson and P. Kumar, \prd{80}{2009}{083529}.
%
\bibitem{kane} P. Grajek, G. Kane, D. Phalen, A. Pierce and S. Watson, \prd{79}{2009}{043506};
G. Kane, R. Lu and S. Watson, \plb{681}{2009}{151}.
D. Feldman, G. Kane, R. Lu and B. Nelson, \plb{687}{2010}{363}.
%
\bibitem{saxions} 
J. E. Kim, \plb{67}{1991}{3465};
D. H. Lyth, \prd{48}{1993}{4523};
S. Chang and H. B. Kim, \prl{77}{1996}{591};
M. Hashimoto, K. Izawa, M. Yamaguchi and T. Yanagida, \plb{437}{1998}{44};
T. Asaka and M. Yamaguchi, \prd{59}{1999}{125003};
J. Hasenkamp and J. Kersten, \prd{82}{2010}{115029};
H. Baer, S. Kraml, A. Lessa and S. Sekmen, arXiv:1012.3769.
%
\bibitem{blrs2} H. Baer, A. Lessa, S. Rajagopalan and W. Sreethawong, 
to appear.
%
\bibitem{mcdonald} J. McDonald, \prd{43}{1991}{1063};
C. Pallis, \app{21}{2004}{689}.
%
\bibitem{lps} G. Lazarides, C. Panagiotakapolous and Q. Shafi, 
\plb{192}{1987}{323};G. Lazarides, R. Schaefer, D. Seckel and Q. Shafi, 
\npb{346}{1990}{193}.
%
\bibitem{vg2} L. Visinelli and P. Gondolo, \prd{81}{2010}{063508}.
%
\bibitem{admx} 
L. Duffy {\it et al.}, \prl{95}{2005}{091304} and \prd{74}{2006}{012006};
for a review, see S. Asztalos, L. Rosenberg, K. van Bibber, P. Sikivie
and K. Zioutas, \arnps{56}{2006}{293}.
%
\end{thebibliography}
\end{document}